\documentclass[11pt,a4paper]{article}
\pdfoutput=1
\usepackage{amsfonts}
\usepackage{jheppub}
\usepackage{amsmath}
\usepackage{amssymb}
\usepackage{color}
\usepackage{graphicx}%
\usepackage[normalem]{ulem}
\newcommand{\stkout}[1]{\ifmmode\text{\sout{\ensuremath{#1}}}\else\sout{#1}\fi}
\allowdisplaybreaks

\title{Einstein Gravity from Conformal Gravity in 6D}

\author[a]{Giorgos Anastasiou,}
\author[b]{Ignacio J. Araya,}
\author[b]{Rodrigo Olea}

\affiliation[a]{Instituto de F\'isica, Pontificia Universidad Cat\'olica de Valpara\'iso,\\ Casilla 4059, Valpara\'iso, Chile.}
\affiliation[b]{Departamento de Ciencias F\'isicas, Universidad Andres Bello,\\
Sazi\'e 2212, Piso 7, Santiago, Chile.}

\emailAdd{georgios.anastasiou@pucv.cl, araya.quezada.ignacio@gmail.com, rodrigo.olea@unab.cl}

\abstract{We extend Maldacena's argument, namely, obtaining Einstein gravity from  Conformal Gravity, to six dimensional manifolds. The proof relies on a particular combination of  conformal (and topological) invariants, which makes manifest the fact that 6D Conformal Gravity admits an Einstein sector. Then, by taking \textit{generalized} Neumann boundary conditions, the Conformal Gravity action reduces to the renormalized Einstein-AdS action. These restrictions are implied by the vanishing of the traceless Ricci tensor, which is the defining property of any Einstein spacetime. The equivalence between Conformal and Einstein gravity renders trivial the Einstein solutions of 6D Critical Gravity at the bicritical point.}

\begin{document}
\maketitle

\section{Introduction}

Conformal mappings of the metric in asymptotically hyperbolic Einstein (AHE) manifolds
are useful to unveil conformal properties of their asymptotic boundaries. Indeed, the conformal
completion technique introduced by Penrose \cite{Penrose:1962ij,Penrose:1964ge} endows these spacetimes with
the notion of {\it conformal infinity}, where the metric can be only specified up to a divergent factor \cite{Graham:1999jg}.
The emergence of conformal symmetry at the boundary from the asymptotic form of the bulk metric,
had been observed in 3D anti-de Sitter (AdS) gravity by Brown and Henneaux \cite{Brown:1986nw} much earlier than the general proposal in the form of AdS/CFT correspondence \cite{Maldacena:1997re,Aharony:1999ti}.

Conformal invariants enter in the physical observables of the dual field theory (e.g., Weyl anomalies) \cite{Henningson:1998gx}.
Furthermore, conformally covariant tensors constructed with the boundary data $g_{(0)ij}$ appear in the asymptotic expansion of the metric,
by solving iteratively the Einstein field equations \cite{hep-th/0002230}. Therefore, conformal invariance/covariance plays an
essential role in the program of Holographic Renormalization.

On the other hand, the renormalization of the Einstein-AdS gravity action  amounts
to the renormalization of the volume for asymptotically AdS (AAdS) solutions. From a more mathematical standpoint,
the concept of Renormalized Volume has been widely explored in the literature \cite{Anderson2000L2CA,Chang:2005ska,1806.10708}.
In the even-dimensional case, the Renormalized Volume for AHE spacetimes is a conformally invariant entity in the bulk. As a consequence,
it is an appealing idea to try to link it to the existence of conformal structures defined throughout the manifold (and not
only at its boundary). This reasoning suggests that the renormalization of the volume/action might be encoded in a mathematical
object --defined for Einstein spaces-- but that can be embedded in a conformally invariant theory of gravity, i.e., a particular version of Conformal Gravity (CG).

A major step in that direction was taken by Maldacena in Ref.\cite{1105.5632}, where he argues that Einstein gravity can be obtained from CG by demanding a Neumann condition in the expansion of the boundary metric. The resulting metric suitably describes the Einstein sector within the 4th-derivative field equations  coming from the Weyl-squared Lagrangian. An equivalent proof, which relies on the cancellation of higher-derivative modes, was provided in Ref.\cite{1608.07826}.
The advantage of the latter procedure is that --after breaking conformal invariance-- it manifestly shows that the form of the action
is the one of MacDowell-Mansouri \cite{PhysRevLett.38.739}, i.e., the renormalized version of Einstein-AdS gravity \cite{Miskovic:2009bm}.

In this paper, we extend Maldacena's argument, i.e., obtaining Einstein gravity from CG, to the six-dimensional case. The proof is based on a particular version of CG, which admits an Einstein sector. By demanding \textit{generalized} Neumann boundary conditions, then, the Conformal Gravity action reduces to the renormalized Einstein-AdS action.

The paper is organized as follows: We first give a general overview of CG in Section \ref{2} and give a useful definition of Einstein manifolds in terms of the vanishing of the trace-free Ricci tensor Section in \ref{3}. We then proceed to derive the Einstein-AdS action starting from CG by imposing the Einstein condition, interpreting it in the holographic viewpoint as a consequence of satisfying generalized Neumann boundary conditions for the metric, both in 4D and in 6D in Sections \ref{4} and \ref{5} respectively. Then, in Section \ref{6}, we consider a definition of critical gravity in 6D, which illustrates how separating the action into Einstein and non-Einstein modes makes manifest the triviality of the action for Einstein spacetimes. Finally, we give some general discussion and mention future avenues of research in Section \ref{7}.

\section{Conformal Gravity}\label{2}

Conformal Gravity is a theory invariant under local Weyl rescalings of the metric $ g_{\mu \nu} \rightarrow \Omega^2 \left(x\right) g_{\mu \nu}$ and it is defined by a linear combination of the conformal invariants of the corresponding dimension. Early works on the subject can be found in \cite{Tomboulis:1984dd,Julve:1978xn,Fradkin:1981iu}. CG, as a higher derivative theory, has better renormalizability and UV properties than Einstein gravity, although it has modes with negative kinetic energy (ghosts).

The most well-studied example of CG is the 4D case due to the fact that Einstein spaces are solutions of the theory. In this case, the action is given by the square of the Weyl tensor (${\lvert W \rvert}^2$), being this the only conformal invariant in four dimensions. Also, the solutions of the 4D CG are conformally Einstein spaces, such as the general spherically symmetric ones with AdS asymptotics \cite{Riegert:1984zz}, charged rotating black holes \cite{Liu:2012xn} and the generalized Plebanski solutions \cite{Mannheim:1990ya}. CG has been applied extensively, as well, in phenomenological aspects as an alternative to dark matter for the explanation of the galaxy rotation curves \cite{Mannheim:1988dj,Mannheim:2005bfa,Mannheim:2010ti,Mannheim:2011ds}. Most importantly, it has been proved a useful tool in constructing conformal supergravity theories  \cite{Kaku:1978nz,Kaku:1978ea,Bergshoeff:1980is,deWit:1980lyi} which emerge naturally from twistor string theory, as shown in \cite{Berkovits:2004jj}.

Generalizing CG in higher dimensions amounts to the ability of determining the conformal invariants of the respective dimensions. The local conformal invariants have been determined in $D=6$ dimensions \cite{Bonora:1985cq,Deser:1993yx,Erdmenger:1997gy} while a detailed classification appears in $D=8$ \cite{Boulanger:2004zf} as well. Our analysis is focused in $D=6$ CG and the 8D case will be addressed in future work.

Unlike CG in four dimensions where a unique conformal invariant appears, i.e. the Weyl tensor squared, the situation in 6D is more complicated due to the presence of three independent conformal invariants \cite{Bastianelli:2000rs,Metsaev:2010kp,Oliva:2010zd}. Two of them are defined in terms of inequivalent contractions of the Weyl tensor as%
\begin{align}
I_{1} &  =W_{\alpha\beta\mu\nu}W^{\alpha\rho\lambda\nu}W_{\rho
~\ \ \ \lambda}^{~\  \beta\mu},\label{I1}\\
I_{2} &  =W_{\mu\nu\alpha\beta}W^{\alpha\beta\rho\lambda}W_{\rho\lambda
}^{~\  \ \mu\nu},
\label{I2}
\end{align}
whereas the third one acquires a non-trivial form as%
\begin{align}
I_{3} &  =W_{\mu\rho\sigma\lambda}\left(  \delta_{\nu}^{\mu}%
\square+4R_{ \nu}^{\mu}-\frac{6}{5}R\delta_{\nu}^{\mu}\right)  W^{\nu
\rho\sigma\lambda}+\nabla_{\mu} J^{\mu},\label{I3}\\
J_{\mu} &  =4R_{\mu}^{~~\lambda\rho\sigma}\nabla^{\nu}R_{\nu
\lambda\rho\sigma}+3R^{\nu\lambda\rho\sigma}\nabla_{\mu}R_{\nu\lambda
\rho\sigma}-5R^{\nu\lambda}\nabla_{\mu}R_{\nu\lambda}+\frac{1}{2}R\nabla_{\mu
}R-R_{\mu}^{ \nu}\nabla_{\nu}R+2R^{\nu\lambda}\nabla_{\nu}R_{\lambda\mu}.
\label{J}
\end{align}
A linear combination of these conformal invariants defines a three-parameter family of Conformal Gravities in six dimensions, whose Lagrangian obtains the general form  $\mathcal{L}=c_{1}I_{1}+c_{2}I_{2}+c_{3}I_{3}$. Contrary to 4D CG, Einstein spacetimes are not solutions of 6D CG for an arbitrary value of the parameters. Indeed, there is a unique choice of parameters where a Scwarzschild-AdS black hole is admitted in the solutions space of the theory, which reads $c_{1}=4c_{2}=-12c_{3}$ \cite{1301.7083}. This choice in 6D is denoted as Lu, Pang and Pope (LPP) CG, and the fact that it has an Einstein sector despite being a higher derivative theory can be understood based on the corresponding Lagrangian having no explicit dependence on Riemann$^2$ or Riemann$^3$ terms (up to the topological term in 6D).

The same conformal invariants that define the CG actions are considered in the definition of renormalized volume \cite{Albin:2005qka,Chang:2005ska,Graham:1999jg,1501.01308,Anderson2000L2CA}. In particular, the renormalized volume $Vol_{ren} \left(M\right)$ of a four-dimensional manifold $M$ was shown in \cite{Anderson2000L2CA} to be proportional to the 4D CG Lagrangian with a fixed coupling constant

\begin{equation}
\frac{1}{32\pi^{2}}%
{\displaystyle\int\limits_{M_{4}}}
d^{4}x\sqrt{\mathcal{G}}W^{\alpha\beta\mu\nu}W_{\alpha\beta\mu\nu}=\chi\left[
M_{4}\right]  -\frac{3}{4\pi^{2}}Vol_{ren}\left[  M_{4}\right]  ,
\end{equation}
where $\chi \left(M\right)$ is the Euler characteristic of the manifold. The presence of the CG Lagrangian in the renormalized volume expression of a four-dimensional manifolds is suggestive for the deeper relation between the two. We remark that the renormalized volume functional defined in terms of the conformal invariants computes the value of the finite universal part in the asymptotic expansion of the volume for AHE manifolds of even dimensions, as explained by Graham in \cite{Graham:1999jg}. 

In general, the direct computation of this universal part requires an appropriate procedure to cancel the divergences of the volume element in AAdS spaces. Recently, it has been shown that when an AAdS manifold with a conformally flat boundary is considered, the counterterms needed for the regularization of its volume acquire a closed form expression that depends explicitly on the extrinsic curvature of the boundary \cite{Anastasiou:2020zwc}. This resummed form of the counterterms in even-D \cite{Olea:2006vd}, is equivalent to the Euler term of the corresponding dimension, up to the Euler characteristic. That is the reason why this scheme is called {\it{Topological Renormalization}}.

In addition to providing an explicit realization of the CG action and being used in the definition of renormalized volume, conformal invariants are considered in the computation of the type-B anomalies of a CFT and play a prominent role in the context of conformal geometry, as well \cite{AST_1985__S131__95_0,Anderson2000L2CA,Chang:2005ska,Alexakis:2010zz,Albin:2005qka,feng2016volume,Gover_2003,bailey1994,Gover_2017,Gover_2019}.

As the main goal of this work is to obtain the Einstein-AdS action from a version of CG with an Einstein sector, a proper characterization of Einstein manifolds will be crucial in this derivation. Indeed, reducing the metric of a CG solution to the one of the Einstein subclass will allow us to constrain the conformal invariants to their restricted form, making the matching apparent.

\section{Einstein spacetimes}\label{3}
\subsection{Definition}
In this work, it is crucial to seek an object that probes the deviation of a spacetime from the Einstein condition. Knowing that Einstein spaces appear as solutions of the field equations of \thinspace$D$-dimensional GR, we get%
\begin{equation}
R_{\nu}^{\mu}-\frac{1}{2}R\delta_{\nu}^{\mu}+\Lambda\delta_{\nu}^{\mu}=0.
\label{EinsteinEOM}%
\end{equation}
Taking the trace of this equation leads to the formula%
\begin{equation}
R=\frac{2D}{D-2}\Lambda=-\frac{D\left(  D-1\right)  }{\ell^{2}}
\label{EinsteinRicciScalar}%
\end{equation}
which, when replaced in \eqref{EinsteinEOM} gives
\begin{equation}
 R_{\nu}^{\mu}-\frac{1}{D}R\delta_{\nu}^{\mu}=0.
\end{equation}
We thus realize that, for Einstein spacetimes with AdS asymptotics, the traceless part of the Ricci tensor,
\begin{equation}
H_{\mu\nu}=R_{\mu\nu}-\frac{1}{D}R \mathcal{G}_{\mu\nu} ,
\label{TracelessRicci}%
\end{equation}
is identically zero. Actually, the divergencelessness of the Einstein tensor indicates that the vanishing of $H_{\nu}^{\mu}$ is a sufficient condition to define a manifold of the Einstein type. Indeed, one may rewrite the Einstein tensor in terms of the traceless Ricci tensor as
\begin{align}
G_{\mu \nu}&= R_{\mu \nu} - \frac{1}{2} R \mathcal{G}_{\mu\nu} \\
&= H_{\mu \nu} - \frac{D-2}{2D} R \mathcal{G}_{\mu\nu} .
\end{align}
Then, considering that $\nabla^{\nu} G_{\mu \nu}=0$ and demanding that $H_{\mu \nu}=0$ implies that $\partial_{\mu}R=0$, which fixes the Ricci scalar to a constant. The value of the Ricci scalar is then determined by the asymptotic structure of the manifold, what in AAdS spacetimes leads to $R=-\frac{D\left(  D-1\right)  }{\ell^{2}}$.

It is straightforward to show that $H_{\nu}^{\mu}$ is different from zero when we are dealing with theories other than Einstein gravity. For instance, the equation of motion (EOM) of Einstein-Weyl gravity in terms of the traceless part of the Ricci tensor, adopts the form
\begin{equation}
H_{\nu}^{\mu}-\frac{D-2}{2D}\left[ R +\frac{D\left(D-1\right)}{\ell^2} \right]\delta_{\nu}^{\mu}+ \gamma B_{\nu}^{\mu}=0, \label{EinsteinWeylEOM}%
\end{equation}
where $\gamma$ is the relative coupling of the Weyl-squared term and $B_{\nu}^{\mu}$ is the Bach tensor, that reads
\begin{equation}
B_{\mu \nu} = \nabla^{\lambda} C_{\mu \nu \lambda} - S^{\lambda \tau} W_{\lambda \mu \nu \tau}  ,
\end{equation}
where $C_{\mu \nu \lambda}$ and $S_{\mu \nu}$ are the Cotton and Schouten tensors defined by
\begin{equation}
C_{\mu\nu\lambda} =\nabla_{\lambda}S_{\mu\nu}- \nabla_{\nu}S_{\mu\lambda}
\label{Cotton}
\end{equation}
and
\begin{equation}
S_{\nu}^{\mu}=\frac{1}{D-2}\left[  R_{\nu}^{\mu}-\frac{1}{2(D-1)}R\delta_{\nu}^{\mu}\right]
\label{Schouten}
\end{equation}
respectively. Also, $W_{\lambda \mu \nu \tau}$ is the Weyl tensor, which is defined in terms of the Schouten tensor as
\begin{equation}
W_{\mu\nu}^{\alpha\beta}=R_{\mu\nu}^{\alpha\beta}-\left(  S_{\mu}^{\alpha
}\delta_{\nu}^{\beta}-S_{\mu}^{\beta}\delta_{\nu}^{\alpha}-S_{\nu}^{\alpha
}\delta_{\mu}^{\beta}+S_{\nu}^{\beta}\delta_{\mu}^{\alpha}\right)  .
\label{Weyltensor}
\end{equation}
Note that the solutions of the theory satisfy the relation $H_{\nu}^{\mu}=-\gamma B_{\nu}^{\mu}$, which corresponds to a broader range of spacetimes than the Einstein ones. The previous analysis indicates that indeed the trace-free Ricci tensor is a sensible way to measure the deviation from Einstein spacetimes.

One may express the Weyl tensor in terms of the trace and the trace-free part of the Ricci tensor. In doing so, the Schouten tensor acquires the form
\begin{equation}
S_{\nu}^{\mu}=\frac{1}{D-2}\left[  H_{\nu}^{\mu}+\frac
{D-2}{2D\left(  D-1\right)  }R\delta_{\nu}^{\mu}\right],
\label{SchoutenTF}
\end{equation}
when said decomposition is applied. It is straightforward to show that, when evaluated on an Einstein spacetime

\begin{equation}
S^{\mu}_{\nu}=-\frac{1}{2\ell^2} \delta^{\mu}_{\nu},
\label{SchoutenEinstein}
\end{equation}
what, in turn, leads to a vanishing Cotton tensor.

As for the Weyl tensor, substituting Eq.\eqref{SchoutenTF} in \eqref{Weyltensor}, one gets

\begin{equation}
W_{\mu\nu}^{\alpha\beta}=R_{\mu\nu}^{\alpha\beta}-\frac{4}{D-2}H_{[\mu}^{[\alpha}\delta_{\nu]}
^{\beta]}-\frac{1}{D\left(  D-1\right)  }R\delta_{\mu\nu}^{\alpha\beta},
\label{WeyltensorTF}
\end{equation}
conveniently expressed in terms of the Ricci scalar and the trace-free part of the Ricci tensor $H_{\mu \nu}$. For reasons that will be clear later, we add and substract the term $\frac{1}%
{\ell^{2}}\delta_{\mu\nu}^{\alpha\beta}$ in order to construct the $AdS$
curvature tensor

\begin{equation}
\mathcal{F}_{\mu\nu}^{\alpha\beta}=R_{\mu\nu}^{\alpha\beta}+\frac
{1}{\ell^{2}}\delta_{\mu\nu}^{\alpha\beta},
\label{FAdS}
\end{equation}
which measures the deviation of the Riemannian curvature relative to pure AdS (constant negative curvature). Hence, the Weyl tensor now reads%
\begin{equation}
W_{\mu\nu}^{\alpha\beta}=\mathcal{F}_{\mu\nu}^{\alpha\beta}-X_{\mu\nu}^{\alpha\beta}
\label{Weyltensordecomp}%
\end{equation}
where the $X$ tensor is given by
\begin{equation}
X_{\mu\nu}^{\alpha\beta}=\frac{4}{D-2}H_{[\mu}^{[\alpha}\delta_{\nu]}^{\beta
]}+\left[  \frac{1}{\ell^{2}}+\frac{1}{D\left(  D-1\right)  }R\right]
\delta_{\mu\nu}^{\alpha\beta}. \label{Xcontrib}%
\end{equation}
Note that the Weyl tensor coincides with the AdS curvature tensor $\mathcal{F}$ only when the Einstein condition is satisfied. This interesting feature makes manifest the presence of the 4D CG action in the renormalized volume of AHE manifolds.

Indeed, in \cite{1806.10708} it was shown that the topologically-renormalized Einstein-AdS action is proportional to the renormalized volume of this class of manifolds. It can be written as a polynomial $P_{D}\left(  \mathcal{F}\right)  $ on powers of the AdS curvature tensor. In the case of 4D, the $P_{4}\left(\mathcal{F}\right)  $ is simply $\left\vert \mathcal{F}\right\vert ^{2}$, which gives the 4D Einstein-AdS gravity theory in its MacDowell-Mansouri form \cite{MacDowell:1977jt}. However, since the AdS curvature is equal to the Weyl curvature, for Einstein spacetimes, one can replace $P_{4}\left(\mathcal{F}\right)  $ with $\left\vert W\right\vert ^{2}$, what is the 4D CG action.

We remark that in CG there is no characteristic scale and the introduction of the AdS radius $\ell$ is justified because we are restricting to manifolds with AdS asymptotics, which are a subset of the possible solutions of the theory.

\subsection{Holographic conditions}

The aforementioned Einstein condition, which was considered from the point of view of the vanishing of the traceless Ricci tensor, can be also understood from the holographic viewpoint in terms of conditions on the asymptotic expansion of the bulk metric. In particular, in the case of manifolds endowed with a negative cosmological constant, the metric acquires an overall double pole that diverges at spacelike infinity. The regularity of the metric can be restored once the Weyl rescaling is applied. This procedure indicates that the pole induces a conformal structure at infinity, and the boundary of the manifold is now a conformal boundary. Fefferman and Graham (FG) in Ref.\cite{AST_1985__S131__95_0}, showed that any AAdS manifold can be written as a power series expansion around the conformal boundary as

\begin{gather}
ds^{2}=\frac{\ell^{2}}{z^{2}}\left(  dz^{2}+g_{ij}\left(  z,x\right)
dx^{i}dx^{j}\right)  \;\;,\label{radialfoliation}\\
g_{ij}\left(  z,x\right)  =g_{\left(  0\right)  ij}\left(  x\right)  +\frac
{z}{\ell}g_{\left(  1\right)  ij}\left(  x\right)  +\frac{z^{2}}{\ell^{2}%
}g_{\left(  2\right)  ij}\left(  x\right)  +\frac{z^{3}}{\ell^{3}}g_{\left(
3\right)  ij}\left(  x\right)  +\ldots,
\label{FGexpansion}
\end{gather}
where $z$ is the radial coordinate.

In Einstein-AdS gravity, this generic expansion is further simplified by the EOM, such that all the odd-powered terms of the FG series but the normalizable mode $g_{\left(  d\right)  ij}$ are forced to be zero. One can then reverse the argument and show, by considering the FG expansion of $H^{\mu}_{\nu}$, that fixing the coefficients of the metric is equivalent to imposing the Einstein condition of $H^{\mu}_{\nu}=0$.

We now proceed to reexamine CG in 4D, understanding the relation between Einstein-AdS and CG as a consequence of the vanishing of the non-Einstein degrees of freedom, which are encoded in terms of the $H^{\mu}_{\nu}$. We also see how in the FG expansion, this vanishing is translated explicitly in specific boundary conditions for the metric.

\section{4D case revisited: Einstein gravity from Conformal Gravity}\label{4}

In \cite{1105.5632}, it is argued that imposing Neumann boundary conditions $\partial_{z}g_{ij}=0$ on the FG expansion of the metric for AAdS spacetimes in CG theory allows to recover the wave function of the universe or the semi-classical partition function for Einstein-AdS gravity. Later in \cite{1608.07826}, the authors arrived at the same conclusion, based on the CG action decomposition into an Einstein and non-Einstein part. Eliminating the latter, they recovered the MacDowell-Mansouri action, what is an equivalent form of the renormalized Einstein-AdS action. In the same spirit, Alaee and Woolgar \cite{1809.06338} determined the asymptotic expansion of Bach flat metrics in CG and provided the appropriate conditions needed to select the Einstein sector of the theory.

Here we reinterpret the construction introduced in \cite{1608.07826}, for generic Bach-flat solutions. Based on the considerations of the previous section, we start from the CG action with an
overall factor $\alpha$
\begin{gather}
I_{CG}=\alpha{\displaystyle\int\limits_{M}}d^{4}x\sqrt{-\mathcal{G}}W_{\mu\nu}^{\alpha\beta}%
W_{\alpha\beta}^{\mu\nu}\\
=\frac{\alpha}{4}{\displaystyle\int\limits_{M}} d^{4}x\sqrt{-\mathcal{G}}\delta_{\nu_{1}\ldots\nu_{4}%
}^{\mu_{1}\ldots\mu_{4}}W_{\mu_{1}\mu_{2}}^{\nu_{1}\nu_{2}}W_{\mu_{3}\mu_{4}%
}^{\nu_{3}\nu_{4}}\label{4DCGaction}%
\end{gather}
and substitute the Weyl tensor in the form given in Eqs.
\eqref{Weyltensordecomp} and \eqref{Xcontrib}. In this case, one gets
\begin{gather}
I_{CG}=\frac{\alpha}{4}{\displaystyle\int\limits_{M}}d^{4}x\sqrt{-\mathcal{G}}\delta_{\nu_{1}\ldots
\nu_{4}}^{\mu_{1}\ldots\mu_{4}}W_{\mu_{1}\mu_{2}}^{\nu_{1}\nu_{2}}\left(
\mathcal{F}_{\mu_{3}\mu_{4}}^{\nu_{3}\nu_{4}}-X_{\mu_{3}\mu_{4}}^{\nu_{3}\nu_{4}%
}\right)  \nonumber\\
=\frac{\alpha}{4}{\displaystyle\int\limits_{M}}d^{4}x\sqrt{-\mathcal{G}}\delta_{\nu_{1}\ldots\nu_{4}%
}^{\mu_{1}\ldots\mu_{4}}W_{\mu_{1}\mu_{2}}^{\nu_{1}\nu_{2}}\mathcal{F}_{\mu_{3}\mu_{4}%
}^{\nu_{3}\nu_{4}}%
\end{gather}
where the term proportional to $X$ vanishes due to the Weyl tensor being traceless. The latter expression can be further accommodated as follows
\begin{gather}
\mathcal{L}_{CG}=\frac{\alpha}{4}\delta_{\nu_{1}\ldots\nu_{4}}^{\mu
_{1}\ldots\mu_{4}}\left(  \mathcal{F}_{\mu_{1}\mu_{2}}^{\nu_{1}\nu_{2}}\mathcal{F}_{\mu_{3}\mu
_{4}}^{\nu_{3}\nu_{4}}-\mathcal{F}_{\mu_{1}\mu_{2}}^{\nu_{1}\nu_{2}}X_{\mu_{3}\mu_{4}%
}^{\nu_{3}\nu_{4}}\right)  \nonumber\\
=\frac{\alpha}{4}\left[  \delta_{\nu_{1}\ldots\nu_{4}}^{\mu_{1}\ldots
\mu_{4}}\mathcal{F}_{\mu_{1}\mu_{2}}^{\nu_{1}\nu_{2}}\mathcal{F}_{\mu_{3}\mu_{4}}^{\nu_{3}\nu_{4}%
}-2\delta_{\nu_{1}\nu_{2}\nu_{3}}^{\mu_{1}\mu_{2}\mu_{3}}\mathcal{F}_{\mu_{1}\mu_{2}%
}^{\nu_{1}\nu_{2}}H_{\mu_{3}}^{\nu_{3}}-\frac{2}{3}\left(  R+\frac{12}%
{\ell^{2}}\right)^{2}\right]
\label{CGactiongeneric}%
\end{gather}
This is a generic decomposition of the CG action for spacetimes with either dS or AdS asymptotics. Here we restrict ourselves to the latter case.

Following the analysis of Section \ref{3}, the presence of the trace-free Ricci tensor in the CG action \eqref{CGactiongeneric} makes expicit the non-Einstein modes of the theory. The different dynamics of CG with respect to Einstein gravity is reflected in a non-vanishing $g_{\left(  1\right)  ij}$ in the FG expansion of Eq.\eqref{FGexpansion}. Nevertheless, due to the fact that the Bach tensor is four-derivative, the EOM does not permit us to dynamically determine coefficients of terms with lower number in derivatives. These modes are actually free data when fixing the formal Poincar\'e-Bach power series \cite{1809.06338}.

One could think that the vanishing of $g_{\left(1\right)  ij}$ is a sufficient condition to select the Einstein sector of the CG, as one recovers the standard FG expansion of AHE metrics. However, a detailed analysis of the asymptotic expansion of the trace-free part of the Ricci tensor $H_{\nu}^{\mu}$, gives
\begin{gather}
H_{z}^{z}=-\frac{z}{4\ell^{3}}g_{\left(  1\right)  }+\frac{z^{2}}{\ell^{2}%
}\left(  \frac{5}{16\ell^{2}}g_{\left(  1\right)  ij}g_{\left(  1\right)
}^{ij}+\frac{1}{16\ell^{2}}g_{\left(  1\right)  }^{2}-\frac{1}{\ell^{2}%
}g_{\left(  2\right)  }-\frac{1}{4}\mathcal{R}_{\left(  0\right)  }\right)
\nonumber\\
+\frac{z^{3}}{\ell^{3}}\left(  -\frac{3}{8\ell^{2}}g_{\left(  1\right)  j}%
^{i}g_{\left(  1\right)  i}^{s}g_{\left(  1\right)  s}^{j}-\frac{1}{8\ell^{2}%
}g_{\left(  1\right)  j}^{i}g_{\left(  1\right)  i}^{j}g_{\left(  1\right)
}+\frac{3}{2\ell^{2}}g_{\left(  1\right)  j}^{i}g_{\left(  2\right)  i}%
^{j}+\frac{1}{4\ell^{2}}g_{\left(  1\right)  }g_{\left(  2\right)  }\right.
\nonumber\\
\left.  -\frac{9}{4\ell^{2}}g_{\left(  3\right)  }+\frac{1}{4}g_{\left(
1\right)  j}^{i}\mathcal{R}_{\left(  0\right)  i}^{j}-\frac{1}{4}D_{\left(
0\right)  i}D_{\left(  0\right)  j}g_{\left(  1\right)  }^{ij}+\frac{1}%
{4}D_{\left(  0\right)  i}D_{\left(  0\right)  }^{i}g_{\left(  1\right)
}\right)  +\mathcal{O}\left(  z^{4}\right)  ,\label{Hzz}%
\end{gather}
\begin{gather}
H_{j}^{i}=\frac{z}{\ell^{3}}\left(  g_{\left(  1\right)  j}^{i}-\frac{1}%
{4}g_{\left(  1\right)  }\delta_{j}^{i}\right)  +\frac{z^{2}}{\ell^{2}}\left(
\frac{1}{16\ell^{2}}g_{\left(  1\right)  ms}g_{\left(  1\right)  }^{ms}%
\delta_{j}^{i}+\frac{1}{16\ell^{2}}g_{\left(  1\right)  }^{2}\delta_{j}%
^{i}-\frac{1}{4\ell^{2}}g_{\left(  1\right)  j}^{i}g_{\left(  1\right)
}\right.  \nonumber\\
\left.  -\frac{1}{2\ell^{2}}g_{\left(  1\right)  j}^{m}g_{\left(  1\right)
m}^{i}+\frac{1}{\ell^{2}}g_{\left(  2\right)  j}^{i}+\mathcal{R}_{\left(
0\right)  j}^{i}-\frac{1}{4}\mathcal{R}_{\left(  0\right)  }\delta_{j}%
^{i}\right)  \nonumber\\
+\frac{z^{3}}{\ell^{3}}\left(  \frac{1}{8\ell^{2}}g_{\left(  1\right)  b}%
^{m}g_{\left(  1\right)  m}^{s}g_{\left(  1\right)  s}^{b}\delta_{j}^{i}%
-\frac{1}{8\ell^{2}}g_{\left(  1\right)  s}^{m}g_{\left(  1\right)  m}%
^{s}g_{\left(  1\right)  }\delta_{j}^{i}+\frac{1}{4\ell^{2}}g_{\left(
1\right)  s}^{m}g_{\left(  1\right)  m}^{s}g_{\left(  1\right)  j}^{i}%
+\frac{3}{4\ell^{2}}g_{\left(  3\right)  }\delta_{j}^{i}\right.  \nonumber\\
+\frac{1}{4\ell^{2}}g_{\left(  1\right)  s}^{i}g_{\left(  1\right)  j}%
^{s}g_{\left(  1\right)  }-\frac{1}{2\ell^{2}}g_{\left(  1\right)  s}%
^{m}g_{\left(  2\right)  m}^{s}\delta_{j}^{i}-\frac{1}{2\ell^{2}}g_{\left(
1\right)  j}^{i}g_{\left(  2\right)  }+\frac{1}{4\ell^{2}}g_{\left(  2\right)
}g_{\left(  1\right)  }\delta_{j}^{i}-\frac{1}{2\ell^{2}}g_{\left(  2\right)
j}^{i}g_{\left(  1\right)  }\nonumber\\
+\frac{1}{4}g_{\left(  1\right)  s}^{m}\mathcal{R}_{\left(  0\right)  m}%
^{s}\delta_{j}^{i}-g_{\left(  1\right)  s}^{i}\mathcal{R}_{\left(  0\right)
j}^{s}-\frac{1}{2}D_{\left(  0\right)  m}D_{\left(  0\right)  }^{m}g_{\left(
1\right)  j}^{i}+\frac{1}{2}D_{\left(  0\right)  m}D_{\left(  0\right)  }%
^{i}g_{\left(  1\right)  j}^{m}+\frac{1}{2}D_{\left(  0\right)  m}D_{\left(
0\right)  j}g_{\left(  1\right)  }^{im}\nonumber\\
\left.  -\frac{1}{4}D_{\left(  0\right)  m}D_{\left(  0\right)  s}g_{\left(
1\right)  }^{ms}\delta_{j}^{i}+\frac{1}{4}\delta_{j}^{i}D_{\left(  0\right)
m}D_{\left(  0\right)  }^{m}g_{\left(  1\right)  }-\frac{1}{2}D_{\left(
0\right)  j}D_{\left(  0\right)  }^{i}g_{\left(  1\right)  }\right)
+\mathcal{O}\left(  z^{4}\right)  ,\label{Hij}%
\end{gather}
and
\begin{gather}
H_{j}^{z}=\frac{z^{2}}{2\ell^{2}}\left(  D_{\left(  0\right)  j}g_{\left(
1\right)  }-D_{\left(  0\right)  m}g_{\left(  1\right)  j}^{m}\right)
+\frac{z^{3}}{\ell^{4}}\left(  -\frac{1}{4}g_{\left(  1\right)  j}%
^{m}D_{\left(  0\right)  m}g_{\left(  1\right)  }+\frac{1}{2}g_{\left(
1\right)  j}^{m}D_{\left(  0\right)  s}g_{\left(  1\right)  m}^{s}\right.
\nonumber\\
\left.  +\frac{1}{2}g_{\left(  1\right)  m}^{s}D_{\left(  0\right)
s}g_{\left(  1\right)  j}^{m}-\frac{3}{4}g_{\left(  1\right)  m}^{s}D_{\left(
0\right)  j}g_{\left(  1\right)  s}^{m}+D_{\left(  0\right)  j}g_{\left(
2\right)  }-D_{\left(  0\right)  m}g_{\left(  2\right)  j}^{m}\right)
+\mathcal{O}\left(  z^{4}\right)  .\label{Hzj}%
\end{gather}
Imposing the condition $g_{\left(  1\right)  ij}=0$ leads to the following expressions for the components of $H_{\nu}^{\mu}$:
\begin{gather}
H_{z}^{z}=-\frac{z^{2}}{\ell^{2}}\left(  \frac{1}{4}\mathcal{R}_{\left(
0\right)  }+\frac{1}{\ell^{2}}g_{\left(  2\right)  }\right)  -\frac{9z^{3}%
}{4\ell^{5}}g_{\left(  3\right)  }+\mathcal{O}\left(  z^{4}\right)
,\label{Hzz2}\\
H_{j}^{i}=\frac{z^{2}}{\ell^{2}}\left(  \frac{1}{\ell^{2}}g_{\left(  2\right)
j}^{i}+\mathcal{R}_{\left(  0\right)  j}^{i}-\frac{1}{4}\mathcal{R}_{\left(
0\right)  }\delta_{j}^{i}\right)  +\frac{3z^{3}}{4\ell^{5}}g_{\left(
3\right)  }\delta_{j}^{i}+\mathcal{O}\left(  z^{4}\right)  ,\label{Hij2}\\
H_{j}^{z}=\frac{z^{3}}{\ell^{4}}\left(  D_{\left(  0\right)  j}g_{\left(
2\right)  }-D_{\left(  0\right)  m}g_{\left(  2\right)  j}^{m}\right)
+\mathcal{O}\left(  z^{4}\right)  .\label{Hzj2}%
\end{gather}
As told above, unlike the Einstein gravity case, the coefficients $g_{\left(2\right)  ij}$ and $tr\left(  g_{\left(  3\right)  ij}\right)  =g_{\left(3\right)  }$ remain undetermined by the Bach-flat condition \cite{1809.06338}. However, the Penrose-Brown-Henneaux (PBH) transformations, which are valid for any AAdS spacetime regardless the EOM of the theory, keep the form of Eqs.\eqref{radialfoliation} and \eqref{FGexpansion} invariant and constrain the form of $g_{\left(  2\right)  ij}=-\ell^{2}S_{\left(  0\right)  ij}$ universally\footnote{For gravity theories with a degenerate AdS branch the situation is different, as discussed in \cite{hep-th/0404245}.} \cite{hep-th/9910267}.

As a consequence, the Neumann boundary condition $\partial_{z}g_{ij}\left(x\right)  |_{z=0}=0$ for Bach-flat solutions with AAdS asymptotics, which sets $g_{\left(  1\right)  ij}=0$, together with the assumption of $g_{\left(  2\right)  ij}=-\ell^{2}S_{\left(  0\right)  ij}$ as given by the PBH transformation, and $g_{\left( 3\right)  }=0$, lead to the vanishing of $H_{\nu}^{\mu}$ up to the normalizable order.

In turn, as stated in the last section, sending $H_{\nu}^{\mu} = 0$ uniquely determines the value of the Ricci scalar, such that the CG action \eqref{CGactiongeneric} reduces to

\begin{equation}
I_{CG} \left[E\right]=\frac{\alpha_{\text{E}}}{4} {\displaystyle\int\limits_{M}}d^{4}x\sqrt{-\mathcal{G}}\delta_{\nu
_{1}\ldots\nu_{4}}^{\mu_{1}\ldots\mu_{4}}\mathcal{F}_{\mu_{1}\mu_{2}}^{\nu_{1}\nu_{2}%
}\mathcal{F}_{\mu_{3}\mu_{4}}^{\nu_{3}\nu_{4}}, \label{4DCGEinstein}
\end{equation}
what is the MacDowell-Mansouri action. Note, that this expression matches the renormalized Einstein-AdS gravity action \cite{1806.10708} 
\begin{equation}
I_{EH}^{ren}=\frac{\ell^{2}}{256\pi G}{\displaystyle\int\limits_{M}}d^{4}x\sqrt{-\mathcal{G}}\delta_{\nu
_{1}\ldots\nu_{4}}^{\mu_{1}\ldots\mu_{4}}\mathcal{F}_{\mu_{1}\mu_{2}}^{\nu_{1}\nu_{2}%
}\mathcal{F}_{\mu_{3}\mu_{4}}^{\nu_{3}\nu_{4}}-\frac{\pi\ell^{2}}{2G}\chi\left(
M\right)  ,
\label{reneinsteinads}%
\end{equation}
up to a topological number\footnote{The Euler characteristic does not play any role in the dynamics of the theory, however it relates topological properties with holographic entanglement entropy computations \cite{Anastasiou:2018rla}.}, in a unique point in the parametric space of CG coupling constants $\alpha_{\text{E}}=\frac{\ell^{2}}{64\pi G}$.  

As a consequence, the Neumann boundary conditions $\partial_{z}g_{ij}|_{z=0}=0$ and $tr\left(  \partial_{z}^{3}g_{ij}|_{z=0}\right)  =0$, under the assumption that $g_{\left(  2\right) ij}=-\ell^{2}S_{\left(  0\right)  ij}$ as given by the PBH analysis, are sufficient to select the Einstein branch out of the AAdS Bach-flat solutions of the theory. The equivalence between CG and renormalized Einstein-AdS gravity can be explicitly realized only when the coupling is fixed as $\alpha_{\text{E}}=\frac{\ell^{2}}{64\pi G}$, on top of the previous conditions.

Here, the additional constraint on $g_{\left(  3\right)  }$ allows us to eliminate  potentially finite contributions to the two terms other than MacDowell-Mansouri in Eq.\eqref{CGactiongeneric}, which become relevant at the level of the on-shell action or the charges, leading to a mismatch between the two theories.

A characteristic example is provided by the computation of the holographic stress-energy tensor of the theory. In this case, the variation of the CG action reads
\begin{equation}
\delta I_{CG}=\alpha{\displaystyle\int\limits_{\partial M}}d^{3}x\sqrt{-g_{\left(  0\right)  }%
}\left[  T_{j}^{i}\left(  g_{(0)}^{-1}\delta g_{\left(  0\right)  }\right)
_{i}^{j}+\tau_{j}^{i}\left(  g_{\left(  0\right)  }^{-1}\delta g_{\left(
1\right)  }\right)  _{i}^{j}\right]  , \label{holovari}%
\end{equation}
where the response function $\tau_{j}^{i}$ is given by 
\begin{equation}
\tau_{j}^{i}=\frac{1}{\ell^{5}}\left[  2g_{\left(  2\right)  j}^{i}+\frac
{1}{6}g_{\left(  1\right)  }^{2}\delta_{j}^{i}-\frac{1}{2}g_{\left(  1\right)
}g_{\left(  1\right)  j}^{i}-\frac{2}{3}g_{\left(  2\right)  }\delta_{j}%
^{i}-\frac{2\ell^{2}}{3}\mathcal{R}_{\left(  0\right)  }\delta_{j}^{i}%
+2\ell^{2}\mathcal{R}_{\left(  0\right)  j}^{i}\right]  ,
\end{equation}
and the holographic stress tensor $T_{j}^{i}$ is defined as 
\begin{gather}
T_{j}^{i}=\frac{1}{\ell^{6}}\left[  \left(  \frac{1}{2}g_{\left(  1\right)
b}^{a}g_{\left(  1\right)  c}^{b}g_{\left(  1\right)  a}^{c}-\frac{1}%
{6}g_{\left(  1\right)  }g_{\left(  1\right)  b}^{a}g_{\left(  1\right)
a}^{b}-\frac{5}{3}g_{\left(  1\right)  b}^{a}g_{\left(  2\right)  a}%
^{b}\right.  \right. \nonumber\\
\left.  +\frac{1}{3}g_{\left(  1\right)  }g_{\left(  2\right)  }+2g_{\left(
3\right)  }+\frac{\ell^{2}}{3}g_{\left(  1\right)  b}^{a}\mathcal{R}_{\left(
0\right)  b}^{a}-\frac{4\ell^{2}}{3}D_{\left(  0\right)  a}D_{\left(
0\right)  b}g_{\left(  1\right)  }^{ba}+\frac{4\ell^{2}}{3}\square_{\left(
0\right)  }g_{\left(  1\right)  }\right)  \delta_{j}^{i}\nonumber\\
+\frac{1}{4}g_{\left(  1\right)  b}^{a}g_{\left(  1\right)  a}^{b}g_{\left(
1\right)  j}^{i}-\frac{3}{2}g_{\left(  1\right)  a}^{i}g_{\left(  1\right)
b}^{a}g_{\left(  1\right)  j}^{b}+\frac{1}{12}g_{\left(  1\right)  }%
^{2}g_{\left(  1\right)  j}^{i}-g_{\left(  1\right)  }g_{\left(  2\right)
j}^{i}+4g_{\left(  1\right)  j}^{a}g_{\left(  2\right)  a}^{ia}\nonumber\\
+2g_{\left(  1\right)  a}^{i}g_{\left(  2\right)  j}^{a}-\frac{1}{3}g_{\left(
1\right)  j}^{i}g_{\left(  2\right)  }-6g_{\left(  3\right)  j}^{i}-2\ell
^{2}g_{\left(  1\right)  a}^{i}\mathcal{R}_{\left(  0\right)  j}^{a}-\frac
{1}{3}g_{\left(  1\right)  j}^{i}\mathcal{R}_{\left(  0\right)  }\nonumber\\
+2\ell^{2}g_{\left(  1\right)  b}^{a}\mathcal{R}_{\left(  0\right)  bj}%
^{ai}+\ell^{2}D_{\left(  0\right)  }^{i}D_{\left(  0\right)  a}g_{\left(
1\right)  j}^{a}-\ell^{2}D_{\left(  0\right)  j}D_{\left(  0\right)  }%
^{i}g_{\left(  1\right)  }-3\ell^{2}D_{\left(  0\right)  j}D_{\left(
0\right)  a}g_{\left(  1\right)  }^{ai}\nonumber\\
\left.  -\ell^{2}D_{\left(  0\right)  a}D_{\left(  0\right)  }^{i}g_{\left(
1\right)  j}^{a}+3\ell^{2}D_{\left(  0\right)  a}D_{\left(  0\right)
j}g_{\left(  1\right)  }^{ai}+\ell^{2}\square_{\left(  0\right)  }g_{\left(
1\right)  j}^{i}\right]  . \label{Tij}%
\end{gather}
The variation in \eqref{holovari} indicates the presence of two independent fields, which are the sources of the dual CFT. They correspond to the holographic duals of two propagating modes, namely $g_{\left(  0\right)  ij}$ for the massless graviton and $g_{\left(  1\right)  ij}$ for the partially massless graviton introduced in CG. Our result matches the formula provided by
Grumiller et al. in \cite{1310.0819}. Imposing the boundary condition
$\partial_{z}g_{ij}|_{z=0}=g_{\left(  1\right)  ij}=0$ while fixing the CG coupling constant to $\alpha_{\text{E}}=\frac{\ell^{2}}{64\pi G}$ leads to the vanishing of the partially massless response function $\tau_{j}^{i}$ associated to $g_{(1)ij}$, whereas the quasilocal stress-energy tensor, which corresponds to the response function associated with $g_{(0)ij}$, reads%
\begin{equation}
T_{j}^{i}=-\frac{3}{32\pi G}\left(  g_{\left(  3\right)  j}^{i}-\frac{1}%
{3}g_{\left(  3\right)  }\delta_{j}^{i}\right)  , \label{TijCG}%
\end{equation}
in accordance with \cite{1310.0819}. Based on this, one may calculate the holographic stress-energy tensor as 
\begin{equation}
\left\langle T^{ij}\right\rangle =-\frac{2}{\sqrt{-g_{\left(  0\right)  }}
}\frac{\delta I_{CG}}{\delta g_{\left(  0\right) ij  }}=\frac{3\ell^{2}}{16\pi G}\left(  g_{\left(  3\right)}^{ij}-\frac{1}{3}g_{\left(  3\right)
}g_{\left(  0\right) }^{ij}\right).
\label{HSETCG}
\end{equation}
Thus, applying the constraint $g_{\left(  3\right) }=0$, Eq.\eqref{HSETCG} reduces to the holographic energy-momentum tensor in Einstein-AdS gravity, given in \cite{hep-th/0002230}.

\section{Renormalized Einstein-AdS action from 6D Conformal Gravity}\label{5}

\subsection{Conformal Gravity with an Einstein sector}

We now begin our analysis of the 6D case by considering the unique combination of the conformal invariants $I_{1},I_{2},I_{3}$, given in Eqs.\eqref{I1},  \eqref{I2} and \eqref{I3}, that admits Einstein solutions
when taken as a Lagrangian for a gravity theory \citep{1301.7083}. As noted in Section \ref{2}, that particular choice is
given by
\begin{equation}
L_{CG}=4I_{1}+I_{2}-\frac{1}{3} I_{3}.
\label{LCGpang}
\end{equation}
We then rewrite the above expression as
\begin{equation}
L_{CG}=\frac{4}{3}\left(  2I_{1}+I_{2}\right)  +\frac{1}{3}\left(
4I_{1}-I_{2}\right)  -\frac{1}{3} I_{3}.
\end{equation}
The first term adopts a compact form, as a polynomial of the Weyl tensor, that reads
\begin{equation}
\delta_{\mu_{1}\cdots\mu_{6}}^{\nu_{1}\cdots\nu_{6}}W_{\nu_{1}\nu_{2}}%
^{\mu_{1}\mu_{2}}W_{\nu_{3}\nu_{4}}^{\mu_{3}\mu_{4}}W_{\nu_{5}\nu_{6}}%
^{\mu_{5}\mu_{6}}=32\left(  2I_{1}+I_{2}\right)  .
\end{equation}
The latter form, along with the identity
\begin{equation}
4I_{1}-I_{2}=W_{\mu\nu}^{\kappa\lambda}\nabla^{2}W_{\kappa\lambda}^{\mu\nu
}-8S_{\rho}^{\omega}W_{\nu\lambda}^{\rho\mu}W_{\omega\mu}^{\nu\lambda
}-2SW_{\nu\lambda}^{\rho\mu}W_{\rho\mu}^{\nu\lambda}+24C^{\mu\nu\lambda}%
C_{\mu\nu\lambda}+8\nabla_{\omega}\left(  W^{\omega\mu\nu\lambda}C_{\mu
\nu\lambda}\right)  ,
\end{equation}
given by Osborn and Stergiou in \cite{1501.01308}, allow us to rewrite \eqref{LCGpang} in the following form,
\begin{align}
L_{CG} &  =\frac{1}{4!}\delta_{\mu_{1}\cdots\mu_{6}}^{\nu_{1}\cdots\nu_{6}%
}W_{\nu_{1}\nu_{2}}^{\mu_{1}\mu_{2}}W_{\nu_{3}\nu_{4}}^{\mu_{3}\mu_{4}}%
W_{\nu_{5}\nu_{6}}^{\mu_{5}\mu_{6}}\nonumber\\
&  +\frac{1}{3}\left(  W_{\mu\nu}^{\kappa\lambda}\nabla^{2}W_{\kappa\lambda
}^{\mu\nu}-8S_{\rho}^{\omega}W_{\nu\lambda}^{\rho\mu}W_{\omega\mu}^{\nu
\lambda}-2SW_{\nu\lambda}^{\rho\mu}W_{\rho\mu}^{\nu\lambda}+24C^{\mu\nu
\lambda}C_{\mu\nu\lambda}+8\nabla_{\omega}\left(  W^{\omega\mu\nu\lambda
}C_{\mu\nu\lambda}\right)  \right)  \nonumber\\
&  -\frac{1}{3}W_{\mu\nu}^{\kappa\lambda}\nabla^{2}W_{\kappa\lambda}^{\mu\nu
}-\frac{4}{3}R_{\rho}^{\omega}W_{\nu\lambda}^{\rho\mu}W_{\omega\mu}%
^{\nu\lambda}+\frac{2}{5}RW_{\nu\lambda}^{\omega\mu}W_{\omega\mu}^{\nu\lambda
}-\frac{1}{3}\nabla_{\mu} J^{\mu}.
\end{align}
Now, using Eq.\eqref{Schouten} we can further simplify $L_{CG}$ into
\begin{align}
L_{CG}  &  =\frac{1}{4!}\delta_{\mu_{1}\cdots\mu_{6}}^{\nu_{1}\cdots\nu_{6}%
}W_{\nu_{1}\nu_{2}}^{\mu_{1}\mu_{2}}W_{\nu_{3}\nu_{4}}^{\mu_{3}\mu_{4}}%
W_{\nu_{5}\nu_{6}}^{\mu_{5}\mu_{6}}-2R_{\rho}^{\omega}W_{\nu\lambda}^{\rho\mu
}W_{\omega\mu}^{\nu\lambda}+\frac{2}{5}RW_{\nu\lambda}^{\omega\mu}W_{\omega
\mu}^{\nu\lambda}+8C^{\mu\nu\lambda}C_{\mu\nu\lambda}\nonumber\\
&  +\frac{1}{3}\nabla_{\mu}\left(  8W^{\mu\kappa\nu\lambda}C_{\kappa\nu
\lambda}- J^{\mu}\right)  .
\end{align}
The CG Lagrangian can be conveniently rewritten once the following identity is considered: 
\begin{equation}
\delta_{\mu_{1}\cdots\mu_{5}}^{\nu_{1}\cdots\nu_{5}}W_{\nu_{1}\nu_{2}}%
^{\mu_{1}\mu_{2}}W_{\nu_{3}\nu_{4}}^{\mu_{3}\mu_{4}}S_{\nu_{5}}^{\mu_{5}} = -4R_{\rho}^{\omega}W_{\nu\lambda}^{\rho\mu}W_{\omega\mu}^{\nu\lambda
}+\frac{4}{5}RW_{\nu\lambda}^{\omega\mu}W_{\omega\mu}^{\nu\lambda}.
\end{equation}
Then, $L_{CG}$ becomes 
\begin{align}
L_{CG}  &  =\frac{1}{4!}\delta_{\mu_{1}\cdots\mu_{6}}^{\nu_{1}\cdots\nu_{6}%
}W_{\nu_{1}\nu_{2}}^{\mu_{1}\mu_{2}}W_{\nu_{3}\nu_{4}}^{\mu_{3}\mu_{4}}%
W_{\nu_{5}\nu_{6}}^{\mu_{5}\mu_{6}}+\frac{1}{2}\delta_{\mu_{1}\cdots\mu_{5}%
}^{\nu_{1}\cdots\nu_{5}}W_{\nu_{1}\nu_{2}}^{\mu_{1}\mu_{2}}W_{\nu_{3}\nu_{4}%
}^{\mu_{3}\mu_{4}}S_{\nu_{5}}^{\mu_{5}}\nonumber\\
&  +8C^{\mu\nu\lambda}C_{\mu\nu\lambda}+\frac{1}{3}\nabla_{\mu}\left(
8W^{\mu\kappa\nu\lambda}C_{\kappa\nu\lambda}-J^{\mu}\right)  .
\label{LCGbulk1}
\end{align}
In the last part of this analysis, we  proceed to simplify the total derivative below. Indeed, considering Eq.\eqref{J}, the total derivative term $T$ of Eq.\eqref{LCGbulk1} is given by
\begin{equation}
T  =\frac{1}{3}\nabla_{\mu}\left(  8W^{\mu\kappa\nu\lambda}C_{\kappa\nu
\lambda}-4R_{\rho\sigma}^{\mu\lambda}\nabla^{\nu}R_{\nu\lambda}^{\rho\sigma
}-3R_{\rho\sigma}^{\nu\lambda}\nabla^{\mu}R_{\nu\lambda}^{\rho\sigma
}+5R_{\lambda}^{\nu}\nabla^{\mu}R_{\nu}^{\lambda}-\frac{1}{2}R\nabla^{\mu
}R+R_{\nu}^{\mu}\nabla^{\nu}R-2R_{\nu}^{\lambda}\nabla^{\nu}R_{\lambda}^{\mu
}\right).
\end{equation}
This expression for $T$ may be further simplified through integration by parts, assuming a smooth boundary manifold, and by considering the relation between the Cotton and Weyl tensors given by
\begin{equation}
C_{\nu\rho\lambda}  =-\frac{1}{\left(  D-3\right)  }\nabla_{\mu}%
W_{~~\nu\rho\lambda}^{\mu}.
\end{equation}
This allows to write $T$ compactly as
\begin{equation}
T=\nabla_{\mu}\left(  8W^{\mu\kappa\lambda\nu}C_{\kappa\lambda\nu}%
-W_{\nu\sigma}^{\kappa\lambda}\nabla_{\mu}W_{\kappa\lambda}^{\nu\sigma
}\right)  .
\end{equation}
Therefore, we have that the full $L_{CG}$ Lagrangian takes the form
\begin{align}
L_{CG}  &  =\frac{1}{4!}\delta_{\mu_{1}\cdots\mu_{6}}^{\nu_{1}\cdots\nu_{6}%
}W_{\nu_{1}\nu_{2}}^{\mu_{1}\mu_{2}}W_{\nu_{3}\nu_{4}}^{\mu_{3}\mu_{4}}%
W_{\nu_{5}\nu_{6}}^{\mu_{5}\mu_{6}}+\frac{1}{2}\delta_{\mu_{1}\cdots\mu_{5}%
}^{\nu_{1}\cdots\nu_{5}}W_{\nu_{1}\nu_{2}}^{\mu_{1}\mu_{2}}W_{\nu_{3}\nu_{4}%
}^{\mu_{3}\mu_{4}}S_{\nu_{5}}^{\mu_{5}}+8C^{\mu\nu\lambda}C_{\mu\nu\lambda
}\nonumber\\
&  +\nabla_{\mu}\left(  8W^{\mu\kappa\lambda\nu}C_{\kappa\lambda\nu}%
-W_{\nu\sigma}^{\kappa\lambda}\nabla^{\mu}W_{\kappa\lambda}^{\nu\sigma
}\right)  ,
\end{align}
and thus, the action is given by
\begin{align}
I_{CG}  &  = \alpha
{\displaystyle\int\limits_{M}}
d^{6}x\sqrt{-\mathcal{G}}\left(  \frac{1}{4!}\delta_{\mu_{1}\cdots\mu_{6}}^{\nu
_{1}\cdots\nu_{6}}W_{\nu_{1}\nu_{2}}^{\mu_{1}\mu_{2}}W_{\nu_{3}\nu_{4}}%
^{\mu_{3}\mu_{4}}W_{\nu_{5}\nu_{6}}^{\mu_{5}\mu_{6}}+\frac{1}{2}\delta
_{\mu_{1}\cdots\mu_{5}}^{\nu_{1}\cdots\nu_{5}}W_{\nu_{1}\nu_{2}}^{\mu_{1}%
\mu_{2}}W_{\nu_{3}\nu_{4}}^{\mu_{3}\mu_{4}}S_{\nu_{5}}^{\mu_{5}}+8C^{\mu
\nu\lambda}C_{\mu\nu\lambda}\right) \nonumber\\
&  + \alpha
{\displaystyle\int\limits_{\partial M}}
d^{5}x\sqrt{-h}n_{\mu}\left(  8W^{\mu \kappa\lambda\nu}C_{\kappa\lambda\nu
}-W_{\nu\sigma}^{\kappa\lambda}\nabla^{\mu}W_{\kappa\lambda}^{\nu\sigma
}\right)  , 
\label{6DCGaction}
\end{align}
where $n_{\mu}$ is the outward normal vector of the foliation and $\alpha$ is an arbitrary overall normalization.

The EOM of the LPP CG action is given by the vanishing of the obstruction tensor $E_{\nu }^{\mu }$, whose explicit form is
\begin{gather}
E_{\nu }^{\mu } = -2\delta _{\nu _{1}\ldots \nu _{4}\nu }^{\mu _{1}\ldots \mu _{4}\mu}W_{\mu _{1}\mu _{2}}^{\nu _{1}\nu _{2}}S_{\mu _{3}}^{\nu _{3}}S_{\mu _{4}}^{\nu _{4}}  
-\frac{1}{2}\delta _{\nu _{1}\ldots \nu _{4}}^{\mu _{1}\ldots \mu _{4}}W_{\mu _{1}\mu _{2}}^{\nu _{1}\nu _{2}}\left ( \nabla ^{\nu _{3}}C_{\mu _{3}\mu _{4}}^{\nu _{4}}\right )\delta _{\nu }^{\mu } \nonumber \\
-\delta _{\nu _{1}\ldots \nu _{4}} ^{\mu _{1}\ldots \mu _{3}\mu }\left [\left ( \nabla ^{\nu _{4}} \nabla _{\nu }W_{\mu _{1}\mu _{2}}^{\nu _{1}\nu _{2}}\right )S_{\mu _{3}}^{\nu _{3}} +\frac{1}{2}\left ( \nabla _{\nu }W_{\mu _{1}\mu _{2}}^{\nu _{1}\nu _{2}}\right )C_{\mu _{3}}^{\nu _{3}\nu _{4}} +W_{\mu _{1}\mu _{2}}^{\nu _{1}\nu _{2}}\left ( \nabla ^{\nu _{4}} \nabla _{\nu }S_{\mu _{3}}^{\nu _{3}}\right )\right ] \nonumber \\
-\delta _{\nu _{1}\ldots \nu _{3}\nu}^{\mu _{1}\ldots \mu _{4}}\left [( \nabla _{\mu _{4}} \nabla ^{\mu }W_{\mu _{1}\mu _{2}}^{\nu _{1}\nu _{2}})S_{\mu _{3}}^{\nu _{3}} +\frac{1}{2}( \nabla ^{\mu }W_{\mu _{1}\mu _{2}}^{\nu _{1}\nu _{2}})C_{\mu _{3}\mu _{4}}^{\nu _{3}} +W_{\mu _{1}\mu _{2}}^{\nu _{1}\nu _{2}}\left ( \nabla _{\mu _{4}} \nabla ^{\mu }S_{\mu _{3}}^{\nu _{3}}\right ) -4W_{\mu _{1}\mu _{2}}^{\nu _{1}\nu _{2}}S_{\mu _{3}}^{\nu _{3}}S_{\mu _{4}}^{\mu }\right ] \nonumber \\
+\delta _{\nu _{1}\ldots \nu _{3}\nu}^{\mu _{1}\ldots \mu _{3}\mu} \Big[ \left ( \nabla ^{\sigma } \nabla _{\sigma }W_{\mu _{1}\mu _{2}}^{\nu _{1}\nu _{2}}\right ) +2\left ( \nabla ^{\sigma }W_{\mu _{1}\mu _{2}}^{\nu _{1}\nu _{2}}\right )\left ( \nabla _{\sigma }S_{\mu _{3}}^{\nu _{3}}\right ) +W_{\mu _{1}\mu _{2}}^{\nu _{1}\nu _{2}}\left ( \nabla ^{\sigma } \nabla _{\sigma }S_{\mu _{3}}^{\nu _{3}}\right ) -  4 C_{\mu _{1}}^{\nu _{1}\nu _{2}}C_{\mu _{2}\mu _{3}}^{\nu _{3}} \nonumber \\
-8 S_{\mu _{1}}^{\nu _{1}}\left ( \nabla ^{\nu _{2}}C_{\mu _{2}\mu _{3}}^{\nu _{3}}\right ) +2 S W_{\mu _{1}\mu _{2}}^{\nu _{1}\nu _{2}}S_{\mu _{3}}^{\nu _{3}} \Big]
+\delta _{\nu _{1}\nu _{2}\nu}^{\mu _{1}\mu _{2}\mu _{3}}\bigg\{ -8 \nabla ^{\nu _{1}}\left (S_{\mu _{1}}^{\mu }C_{\mu _{2}\mu _{3}}^{\nu _{2}}\right ) -4 \nabla _{\mu _{1}}\left (S^{\mu \nu _{1}}C_{\mu _{2}\mu _{3}}^{\nu _{2}}\right ) +14\left ( \nabla ^{\mu }S_{\mu _{1}}^{\nu _{1}}\right )C_{\mu _{2}\mu _{3}}^{\nu _{2}} \nonumber \\ 
+2\left ( \nabla ^{\nu _{1}}C_{\mu _{1}\mu _{2}}^{\nu _{2}} + \nabla _{\mu _{1}}C_{\mu _{2}}^{\nu _{1}\nu _{2}}\right )R_{\mu _{3}}^{\mu }
+4\left [\left ( \nabla _{\mu _{2}}C_{\mu _{1}}^{\nu _{1}\nu _{2}}\right )S_{\mu _{3}}^{\mu } -\frac{1}{2}C_{\mu _{1}}^{\nu _{1}\nu _{2}}C_{\mu _{2}\mu _{3}}^{\mu }\right ] - \nabla _{\mu _{3}} \nabla ^{\mu }\left ( \nabla ^{\nu _{1}}C_{\mu _{1}\mu _{2}}^{\nu _{2}} + \nabla _{\mu _{1}}C_{\mu _{2}}^{\nu _{1}\nu _{2}}\right )\bigg\} \nonumber \\
+\delta _{\nu _{1}\nu _{2}\nu _{3}}^{\mu _{1}\mu _{2}\mu } \Big[ -4 \nabla ^{\nu _{1}}\left (S_{\nu }^{\nu _{2}}C_{\mu _{1}\mu _{2}}^{\nu _{3}}\right ) +4 \nabla ^{\nu _{3}}\left (C_{\mu _{1}}^{\nu _{1}\nu _{2}}S_{\nu \mu _{2}}\right ) +6\left ( \nabla _{\nu }S_{\mu _{1}}^{\nu _{1}}\right )C_{\mu _{2}}^{\nu _{2}\nu _{3}} 
- \nabla ^{\nu _{3}} \nabla _{\nu }\left ( \nabla ^{\nu _{1}}C_{\mu _{1}\mu _{2}}^{\nu _{2}} + \nabla _{\mu _{1}}C_{\mu _{2}}^{\nu _{1}\nu _{2}}\right ) \Big] \nonumber \\
+\delta _{\nu _{1}\nu _{2}\nu}^{\mu _{1}\mu _{2}\mu} \Big[4 \nabla ^{\sigma }\left (S_{\sigma }^{\nu _{1}}C_{\mu _{1}\mu _{2}}^{\nu _{2}}\right ) 
-4 \nabla _{\sigma }\left (C_{\mu _{1}}^{\nu _{1}\nu _{2}}S_{\mu _{2}}^{\sigma }\right ) + \nabla ^{\sigma } \nabla _{\sigma }\left ( \nabla ^{\nu _{1}}C_{\mu _{1}\mu _{2}}^{\nu _{2}} + \nabla _{\mu _{1}}C_{\mu _{2}}^{\nu _{1}\nu _{2}}\right ) \Big] \nonumber \\
+\delta _{\nu _{1}\nu _{2}\nu _{3}}^{\mu _{1}\mu _{2}\mu _{3}} \Big[ \nabla ^{\nu _{3}} \nabla _{\mu _{3}}\left ( \nabla ^{\nu _{1}}C_{\mu _{1}\mu _{2}}^{\nu _{2}} + \nabla _{\mu _{1}}C_{\mu _{2}}^{\nu _{1}\nu _{2}}\right ) +6S_{\mu _{1}}^{\nu _{1}}\left ( \nabla ^{\nu _{2}}C_{\mu _{2}\mu _{3}}^{\nu _{3}}\right ) +8C_{\mu _{1}}^{\nu _{1}\nu _{2}}C_{\mu _{2}\mu _{3}}^{\nu _{3}} \Big] \delta _{\nu }^{\mu } .
\end{gather}
For more details on its derivation see Appendix \ref{EOM}. This tensor vanishes identically for any Einstein spacetime, confirming the fact that the theory has an Einstein sector. In this sense, $E_{\nu }^{\mu }$ is a generalization of the Bach tensor and corresponds to the six-dimensional obstruction tensor.

Furthermore, Eq.\eqref{6DCGaction} provides a compact reformulation of the LPP CG Lagrangian \eqref{LCGpang}. Especially, the presence of the polynomial of the Weyl tensor probes a deeper connection to renormalized volume, where similar structures arise \cite{Albin:2005qka,Chang:2005ska,1806.10708}.

\subsection{Conformal to Einstein gravity from the holographic viewpoint}

The analysis of Section \ref{4} has shown that CG in $D=4$ is classically equivalent to renormalized EH gravity for spacetimes endowed with a negative cosmological constant, when precise boundary conditions are satisfied. In what follows, we extend Maldacena's argument  and determine the boundary conditions required for the LPP action to consistently reduce to the Einstein-AdS action.

In a similar fashion to the 4D case, we apply the Weyl tensor decomposition of Eq.\eqref{Weyltensordecomp} to the action in Eq.\eqref{6DCGaction}. Interestingly enough, the LPP CG action now reads

\begin{gather}
I_{CG}= \alpha{\displaystyle\int\limits_{M}}d^{6}x\sqrt{-\mathcal{G}}\Big[-4! P_{6}\left[ \mathcal{F}\right]+ F_{\text{bulk}} \left[H,R\right]  \Big]  \nonumber\\
+\alpha {\displaystyle\int\limits_{\partial M}}d^{5}x\sqrt{-h}  \Big[-\frac{1}{4}n_{\mu}\delta_{\nu_{1}\ldots\nu_{4}}^{\mu_{1}\ldots\mu_{4}}\mathcal{F}_{\mu_{1}\mu_{2}}^{\nu_{1}\nu_{2}}\nabla^{\mu}\mathcal{F}_{\mu_{3}\mu_{4}}^{\nu_{3}\nu_{4}} + F_{\text{surf}} \left[H,R\right]  \Big]  \,,
\label{6DCGgeneric}
\end{gather}
where
\begin{equation}
P_{6}\left(  \mathcal{F}\right)  =-\frac{1}{\left(  4!\right)  ^{2}}%
\delta_{\mu_{1}\cdots\mu_{6}}^{\nu_{1}\cdots\nu_{6}}\mathcal{F}_{\nu_{1}%
\nu_{2}}^{\mu_{1}\mu_{2}}\mathcal{F}_{\nu_{3}\nu_{4}}^{\mu_{3}\mu_{4}%
}\mathcal{F}_{\nu_{5}\nu_{6}}^{\mu_{5}\mu_{6}}+\frac{1}{2\left(  4!\right)
\ell^{2}}\delta_{\mu_{1}\cdots\mu_{4}}^{\nu_{1}\cdots\nu_{4}}\mathcal{F}%
_{\nu_{1}\nu_{2}}^{\mu_{1}\mu_{2}}\mathcal{F}_{\nu_{3}\nu_{4}}^{\mu_{3}\mu
_{4}}
\label{P6}
\end{equation}
and 
\begin{gather}
F_{\text{bulk}} \left[H,R\right] =  -\frac{1}{4}\delta_{\nu_{1}\ldots\nu
_{4}}^{\mu_{1}\ldots\mu_{4}} \mathcal{F}_{\mu_{1}\mu_{2}}^{\nu_{1}\nu_{2}}H_{\mu_{3}
}^{\nu_{3}}H_{\mu_{4}}^{\nu_{4}}+\frac{1}{2}\delta_{\nu_{1}\nu_{2}\nu_{3}%
}^{\mu_{1}\mu_{2}\mu_{3}}\left(  H_{\mu_{1}}^{\nu_{1}}H_{\mu_{2}}^{\nu_{2}%
}-\frac{1}{5}\mathcal{F}_{\mu_{1}\mu_{2}}^{\nu_{1}\nu_{2}}\right)  H_{\mu_{3}}^{\nu_{3}%
} \nonumber\\
+\frac{2}{5}\left(  \frac{15}{\ell^{2}}+R\right)  \delta_{\nu_{1}\nu_{2}}%
^{\mu_{1}\mu_{2}}H_{\mu_{1}}^{\nu_{1}}H_{\mu_{2}}^{\nu_{2}}+\nabla_{\lambda
}H_{\nu}^{\mu}\nabla^{\lambda}H_{\mu}^{\nu}-\nabla_{\lambda}H_{\nu}^{\mu
}\nabla^{\nu}H_{\mu}^{\lambda}\nonumber\\
  -\frac{1}{5}\nabla_{\mu}H_{\nu}^{\mu}\nabla^{\lambda}H_{\lambda}^{\nu
}-\frac{2}{225}\left(  R-\frac{60}{\ell^{2}}\right)  \left(  \frac{30}%
{\ell^{2}}+R\right)  ^{2} , \label{Fbulk}\\
F_{\text{surf}} \left[H,R\right] = n_{\mu} \left[ \frac{3}{4}\delta
_{\nu_{1}\nu_{2}\nu_{3}}^{\mu_{1}\mu_{2}\mu_{3}}\left( \mathcal{F}_{\mu_{1}\mu_{2}%
}^{\nu_{1}\nu_{2}}\nabla^{\mu}H_{\mu_{3}}^{\nu_{3}}+H_{\mu_{1}}^{\nu_{1}%
}\nabla^{\mu}\mathcal{F}_{\mu_{2}\mu_{3}}^{\nu_{2}\nu_{3}}\right)  +4\mathcal{F}^{\mu\kappa\lambda\nu}\nabla_{\nu}H_{\kappa \lambda} \right. \nonumber\\
 \left. -H^{\mu\lambda}\nabla_{\nu}H_{\lambda}^{\nu}-H_{\kappa}^{\nu}\nabla_{\nu}H^{\mu\kappa}-2H_{\kappa}^{\lambda}\nabla^{\mu}H_{\lambda}^{\kappa}+\frac{16}{15}\left(  \frac{30}{\ell^{2}}+R\right)  \nabla_{\nu}H^{\mu\nu}\right]  .
\label{Fsurf}
\end{gather}
See Appendix \ref{CGdecomposition} for a detailed derivation. Note, that this decomposition makes manifest the contribution of the non-Einstein modes of the theory, encoded in $F_{\text{bulk}} \left[H,R\right]$ and $F_{\text{surf}} \left[H,R\right]$. Switching-off these modes amounts to identifying the appropriate set of boundary conditions that lead to $H_{\mu \nu}=0$, up to the normalizable order.

Since our analysis focuses on manifolds with AdS asymptotics, the bulk metric can be expressed in the FG gauge \eqref{FGexpansion} with
\begin{equation}
g_{ij}\left(  z,x\right)  =g_{\left(  0\right)  ij}\left(  x\right)  +\frac
{z}{\ell}g_{\left(  1\right)  ij}\left(  x\right)  +\frac{z^{2}}{\ell^{2}%
}g_{\left(  2\right)  ij}\left(  x\right)  +\frac{z^{3}}{\ell^{3}}g_{\left(
3\right)  ij}\left(  x\right)  +\frac{z^{4}}{\ell^{4}}g_{\left(  4\right)
ij}\left(  x\right)  +\frac{z^{5}}{\ell^{5}}g_{\left(  5\right)  ij}\left(
x\right)  +\mathcal{O}\left(  z^{6}\right)  .
\label{6DFGexpansion}%
\end{equation}
The different dynamics of the theory with respect to standard Einstein gravity gives rise to two additional terms in the series, i.e., $g_{\left(  1\right)  ij}$ and $g_{\left(  3\right)  ij}$. The additional odd-power contributions in the FG expansion reflect the presence of two propagating massive spin-2 modes of the theory. Due to the fact that the equations of motion are six-derivative, all the terms of the series up to $\mathcal{O} \left(  z^{6}\right)  $ are dynamically undetermined.

In turn, the identification of the Einstein sector of the theory requires the absence of $g_{\left(  1\right)  ij}$ and $g_{\left(  3\right)  ij}$, as the odd-power terms in the FG expansion of Einstein gravity vanish identically. Thus, one could wonder if the generalized Neumann boundary conditions, $\partial_{z}g_{ij}
|_{z=0}=g_{\left(  1\right)  ij}=0$ and $\partial_{z}^{3}g_{ij}|_{z=0}=6 g_{\left(  3\right)  ij}=0$, are sufficient to fix $H_{\nu}^{\mu}=0$ up to the normalizable order. Under these conditions, the asymptotic analysis of the trace-free part of the Ricci tensor $H_{\nu}^{\mu}$ gives that
\begin{gather}
H_{z}^{z}=-\frac{25z^{5}}{3\ell^{7}}g_{\left(  5\right)  }-\frac{z^{2}}%
{\ell^{2}}\left(  \frac{4}{3\ell^{2}}g_{\left(  2\right)  }+\frac{1}%
{6}\mathcal{R}_{\left(  0\right)  }\right)
+\frac{z^{4}}{\ell^{4}}\left(  \frac{11}{6\ell^{2}}g_{\left(  2\right)
ij}g_{\left(  2\right)  }^{ij}+\frac{1}{6\ell^{2}}g_{\left(  2\right)  }^{2} \right. \nonumber\\
\left. -\frac{16}{3\ell^{2}}g_{\left(  4\right)  }+\frac{1}{6}g_{\left(
2\right)  ij}\mathcal{R}_{\left(  0\right)  }^{ij}-\frac{1}{6}D_{\left(
0\right)  i}D_{\left(  0\right)  j}g_{\left(  2\right)  }^{ij}+\frac{1}%
{6}D_{\left(  0\right)  i}D_{\left(  0\right)  }^{i}g_{\left(  2\right)}\right)  + \mathcal{O} \left(z^6\right),\label{Hzz6D}\\
H_{j}^{i}=\frac{z^{2}}{\ell^{2}}\left(  -\frac{1}{3\ell^{2}}g_{\left(
2\right)  }\delta_{j}^{i}+\frac{3}{\ell^{2}}g_{\left(  2\right)  j}%
^{i}+\mathcal{R}_{\left(  0\right)  j}^{i}-\frac{1}{6}\mathcal{R}_{\left(
0\right)  }\delta_{j}^{i}\right) + \frac{z^4}{\ell^4} \left(-\frac{1}{6 \ell^2} g_{\left(  2\right)  b}^{a}g_{\left(  2\right)
a}^{b} \delta_{j}^{i}+ \frac{1}{6 \ell^2} g_{\left(2\right)}^2 \delta_{j}^{i} \right. \nonumber\\
\left. - \frac{1}{\ell^2} g_{\left(2\right)} g_{\left(  2\right)
j}^{i} - \frac{1}{\ell^2} g_{\left(  2\right)  j}^{a}g_{\left(  2\right) a}^{i} + \frac{2}{3 \ell^2} g_{\left(4\right)} \delta_{j}^{i} + \frac{2}{\ell^2} g_{\left(  4\right) j}^{i} + \frac{1}{6} g_{\left(  2\right)  b}^{a} R_{a}^{b} \delta_{j}^{i} - g_{\left(  2\right)  j}^{a} R_{a}^{i} \right. \nonumber\\
\left.   - \frac{1}{2} D_{\left(  0\right)  a}D_{\left(  0\right)  }^{a} g_{\left(  2\right) j}^{i} + \frac{1}{2} D_{\left(0\right)  a} D_{\left(  0\right)}^{i} g_{\left(  2\right) j}^{a} + \frac{1}{2} D_{\left(0\right)  a} D_{\left(  0\right) j} g_{\left(  2\right)}^{i a} - \frac{1}{6} D_{\left(0\right)  a} D_{\left(  0\right) b} g_{\left(  2\right)}^{a b} \delta_{j}^{i} \right. \nonumber\\
\left.  + \frac{1}{6} \delta_{j}^{i} D_{\left(  0\right)  a}D_{\left(  0\right)  }^{a} g_{\left(  2\right)} - \frac{1}{2} D_{\left(0\right)  j} D_{\left(  0\right)}^{i} g_{\left(  2\right)} \right) + \frac{5 z^5}{3 \ell^7} g_{\left(  5\right)} \delta_{j}^{i} + \mathcal{O} \left(z^6\right) ,
\label{Hij6D}\\
H_{j}^{z}=\frac{z^{3}}{\ell^{4}}\left(  D_{\left(  0\right)  j}g_{\left(
2\right)  }-D_{\left(  0\right)  m}g_{\left(  2\right)  j}^{m}\right)
+\frac{z^{5}}{\ell^{6}}\left(  -\frac{1}{2}g_{\left(  2\right)  j}%
^{m}D_{\left(  0\right)  m}g_{\left(  2\right)  }-2D_{\left(  0\right)
m}g_{\left(  4\right)  j}^{m} \right. \nonumber\\
\left. +2D_{\left(  0\right)  j}g_{\left(  4\right)
}+g_{\left(  2\right)  j}^{m}D_{\left(  0\right)  s}g_{\left(  2\right)
m}^{s}+g_{\left(  2\right)  m}^{s}D_{\left(  0\right)  s}g_{\left(  2\right)
j}^{m}-\frac{3}{2}g_{\left(  2\right)  m}^{s}D_{\left(  0\right)  j}g_{\left(
2\right)  s}^{m}\right)+ \mathcal{O} \left(z^6\right) . \label{Hzj6D}%
\end{gather}
At this point, it seems necessary to set additional data, namely $g_{\left(  2\right)  ij}$, $g_{\left(  4\right)  ij}$ and $g_{\left( 5\right)  }$, in order to fulfill the condition stated above.

We consider that both $g_{\left(  2\right)  ij}$ and $g_{\left(
4\right)  ij}$ are partially fixed by the PBH transformations for every diffeomorphism invariant theory with AAdS asymptotics. As shown in \cite{hep-th/9910267}, they can explicitly be written in the form
\begin{gather}
g_{\left(  2\right)  ij}=-\ell^{2}\mathcal{S}_{\left(  0\right)
ij},\label{g2}\\
g_{\left(  4\right)  ij}=c_{1}\ell^{4}\mathcal{W}_{\left(  0\right)  kl}%
^{ms}\mathcal{W}_{\left(  0\right)  ms}^{kl}g_{\left(  0\right)  ij}+c_{2}%
\ell^{4}\mathcal{W}_{\left(  0\right)  iklm}\mathcal{W}_{\left(  0\right)
j}^{\ \ klm}+\frac{\ell^{4}}{4(d-4)}\left[  \left(  d-4\right)  \mathcal{S}%
_{\left(  0\right)  im}\mathcal{S}_{\left(  0\right)  j}^{m}-\mathcal{B}%
_{\left(  0\right)  ij}\right]  , \label{g4}%
\end{gather}
where $\mathcal{B}_{\left(  0\right)  ij}$ is the Bach tensor of the boundary metric $g_{\left(  0\right)  ij}$, and $c_{1},c_{2}$ are theory dependent constants. In Einstein-AdS gravity, one has that $c_{1}=c_{2}=0$ (see \cite{hep-th/0002230}). Since the Lagrangian \eqref{LCGpang} is a 6D CG that admits an Einstein branch of solutions, the constants $c_{1}$ and $c_{2}$ would be uniquely fixed to zero, as they are theory-dependent and not solution-dependent. We will thus consider that the forms of $g_{\left(2\right)  ij}$ and $g_{\left(  4\right)  ij}$ are as given in the previous equations, which corresponds to the relation satisfied by Einstein gravity.

Substituting Eqs.\eqref{g2} and \eqref{g4} with $c_{1}=c_{2}=0$ into Eqs.\eqref{Hzz6D}, \eqref{Hij6D} and \eqref{Hzj6D}, and considering that \citep{1501.01308}
\begin{equation}
\nabla^{\nu}B_{\mu\nu}   =\left(  D-4\right)  S^{\lambda\rho}C_{\lambda\rho\mu},
\end{equation}
we verify that all the components of $H_{\nu}^{\mu}$ but $H_{z}^{z}$ and $H_{j}^{i}$ vanish up to the normalizable order. The latter do not vanish only due to the presence of $g_{\left(  5\right)  }$.

Hence, constraining the 6D CG to the Einstein branch corresponds to the following boundary conditions: i) $\partial_{z}g_{ij}|_{z=0}=0$, ii) $\partial_{z}^{3}g_{ij}|_{z=0}=0$ and iii) $tr\left(  \partial_{z}^{5}g_{ij}\right)|_{z=0}  =0$, under the assumption that $g_{\left(  2\right)  ij}$ and $g_{\left(  4\right)  ij}$ are given according to Eqs.\eqref{g2} and \eqref{g4} with $c_{1}=c_{2}=0$.

Considering these conditions, the 6D CG action \eqref{6DCGgeneric} acquires the form

\begin{equation}
I_{CG} \left[E\right]=-4! \alpha
{\displaystyle\int\limits_{M}}
d^{6}x\sqrt{-\mathcal{G}}P_{6}\left(  \mathcal{F}\right)  - \alpha
{\displaystyle\int\limits_{\partial M}}
d^{5}x\sqrt{-h}n^{\mu}\left(  \mathcal{F}_{\nu\sigma}^{\kappa\lambda}%
\nabla_{\mu}\mathcal{F}_{\kappa\lambda}^{\nu\sigma}\right)  .
\label{ICGE}
\end{equation}
More importantly, when $\alpha =\alpha_{E}=-\frac{\ell^{4}}{384 \pi G_{N}}$, the following relation is valid (See Appendix \ref{EinsteinCG})

\begin{equation}
I_{EH}^{\text{ren}}= I_{CG}\left[E\right]
+\frac{\pi^{2}\ell^{4}}{3G_{N}}\chi\left(  M\right)  ,
\label{Einstein_TopConf}
\end{equation}
where $I_{EH}^{\text{ren}}$ is the renormalized Einstein-AdS action, given by

\begin{equation}
I_{EH}^{\text{ren}}= I_{EH}^{KT}+\frac{\ell^{4}}{384\pi G_{N}}
{\displaystyle\int\limits_{\partial M}}
d^{5}x\sqrt{-h}n^{\mu}\left(  \mathcal{F}_{\nu\sigma}^{\kappa\lambda}%
\nabla_{\mu}\mathcal{F}_{\kappa\lambda}^{\nu\sigma}\right) .
\label{IrenEH}
\end{equation}
Here $I_{EH}^{KT}$ is the Kounterterm-renormalized Einstein-AdS action \cite{Olea:2006vd} that applies to manifolds with conformally flat boundaries \cite{Anastasiou:2020zwc}. The additional surface term in $I_{EH}^{\text{ren}}$ extends the applicability of the renormalization scheme to manifolds with arbitrary boundary geometry (see Appendix \ref{6DCGdivcancel}).
More importantly, Eq.\eqref{Einstein_TopConf} makes manifest the equivalence between $  I_{CG} \left[E\right]$ and the renormalized Einstein-AdS action $I_{EH}^{\text{ren}}$, up to the Euler characteristic.

Therefore, we are able to find the generalization of Maldacena's Neumann boundary conditions to the case of the LPP Conformal Gravity in 6D, which is given by the three conditions stated above. In the next section, we use the decomposed form of the LPP action, as given in Eq.\eqref{6DCGgeneric}, in order to construct a Critical Gravity action in 6D, which has a bicritical point and which is manifestly trivial for Einstein solutions.

\section{Reinterpretation of 6D Critical Gravity}\label{6}

A particular Lagrangian for 6D Critical Gravity can be constructed in the same way as in the 4D case, i.e., by considering the difference between the renormalized Einstein-AdS action and the CG action that admits Einstein solutions. The resulting theory has a trivial action for Einstein spaces at the bicritical point, as shown in \citep{1106.4657}.
The action can be equivalently rewritten as
\begin{equation}
I_{\text{Crit}}   =I_{EH}^{\text{ren}}-I_{CG} \left(\alpha_{\text{E}}\right)-\frac{\pi^{2}\ell^{4}%
}{3G_{N}}\chi\left(  M\right),
\label{I_Crit}%
\end{equation}
what makes manifest its triviality when the spacetime considered is of Einstein class. Indeed, replacing Eq.\eqref{Einstein_TopConf} in the above action leads to $  I_{\text{Crit}}[E]=0$.

This feature is made explicit by the decomposition of the $I_{CG}$ action, into Einstein and non-Einstein terms (see Appendix \ref{CGdecomposition}). Replacing Eq.\eqref{6DCGgeneric} into the Critical Gravity action \eqref{I_Crit} we have that
\begin{gather}
I_{\text{Crit}}= I_{CG}\left[E\right]-I_{CG} \nonumber\\
=-\alpha_{\text{E}}{\displaystyle\int\limits_{ M}}d^{6}x\sqrt{-\mathcal{G}}F_{\text{bulk}} \left[H,R\right] -\alpha_{\text{E}} {\displaystyle\int\limits_{\partial M}} d^{5}x\sqrt{-h} F_{\text{surf}} \left[H,R\right]  \,,
\label{CritGravNew}
\end{gather}
where $F_{\text{bulk}} \left[H,R\right]$ and $F_{\text{surf}} \left[H,R\right]$ are defined in Eqs.\eqref{Fbulk} and \eqref{Fsurf} respectively.
 As $F_{\text{bulk}} \left[H,R\right]$ and $F_{\text{surf}} \left[H,R\right]$ vanish for Einstein spacetimes by construction, it follows that in the Einstein sector, Critical Gravity becomes trivial. The action is a function of the traceless part of the Ricci tensor $H_{\nu}^{\mu}$. In other words the theory is non-trivial only for non-Einstein spaces.

This situation is analogous to the 4D case studied in
\cite{Anastasiou:2017mag}, where one has%
\begin{equation}
I_{\text{Crit}}=I_{EH}^{\text{ren}}-I_{CG}\left(\alpha_{\text{E}}\right)+\frac{\pi \ell^2}{2G_{N}}\chi\left(M\right),
\end{equation}
such that, for Einstein spacetimes, $  I_{\text{Crit}}\left[E\right]=0$.  In the case of 4D Critical Gravity, the corresponding tensor that encodes the non-Einstein modes is the Bach tensor $B_{\mu \nu}$. Thus, the on-shell action can be written as $\left\vert B\right\vert ^{2}$, i.e.,
\begin{equation}
I_{\text{Crit}}=-\frac{\ell^{6}}{512\pi G_{N}}
{\displaystyle\int\limits_{M}}
d^{4}x\sqrt{-\mathcal{G}}B_{\nu}^{\mu}B_{\mu}^{\nu}.
\end{equation}
Nevertheless, there is a clear distinction here that has to be pointed out. 4D Critical Gravity describes a trivial Einstein sector in a theory with no propagating massive modes. On the contrary, its 6D counterpart features the same trivial Einstein branch but in a point of the parametric space where a massive propagating mode exists. Both theories correspond to a critical point of multiplicity 2, but the criticality criterion, namely the absence of massive excitations, is met only in 4D. The latter leads to a difference between the notions of criticality and triviality in higher-curvature gravity theories. The condition of full criticality, i.e., the cancellation of the massive modes, is satisfied at the tri-critical point. However, this point is defined once an extra Weyl squared term is added in the Lagrangian and, therefore, the action on Einstein spacetimes is no longer trivial \cite{1106.4657,Deser:2011xc}.

\section{Discussion and closing remarks}\label{7}

We have shown that the LPP version of Conformal Gravity in 6D is classically equivalent to the Einstein-AdS action for Einstein spaces. In order to make this equivalence manifest, we have extended Maldacena's holographic argument by finding additional conditions on the metric at the conformal boundary, which fully characterize
Einstein spaces. These constrains are codified in the requirement of $H_{\mu\nu}=0$ up to the normalizable order and correspond to generalized Neumann boundary conditions on $g_{ij}\left(  x,z\right)  $ at $z=0$ :
i) $ \partial_{z}g_{ij}\left(  x,z\right) =0$,
ii) $ \partial_{z}^{3}g_{ij}\left(  x,z\right)  =0$
and iii) $tr\left(  \partial_{z}^{5}g_{ij}\left(  x,z\right)\right)  =0$, under the assumption that the FG coefficients $g_{\left(  2\right)  ij}$ and $g_{\left(  4\right)  ij}$ are given by the corresponding universal form of Eqs.\eqref{g2} and \eqref{g4} in the PBH transformations \cite{hep-th/9910267}. When this analysis was applied  to the 4D case, we recovered the usual Neumann boundary condition $  \partial_{z}g_{ij}\left(
x,z\right)  =g_{(1)ij}=0$ discussed in Ref.\cite{1105.5632}. This is a necessary --but not sufficient-- condition, as it requires 
that $tr\left(   \partial_{z}^{3}g_{ij}\left(  x,z\right) \right) =0$,  and the CG free data $g_{\left(  2\right)  ij}$  determined as the Schouten tensor of $g_{(0)}$ by the PBH near-boundary analysis.

We have also shown that any 6D Einstein-AdS manifold is a solution of the LPP CG. To accomplish this, we have derived the equation of motion of CG starting from the compact form of the action presented in Eq.\eqref{LCGpang}, and we have verified that if the Einstein condition of $H^{\mu}_{\nu}=0$ is imposed, then the EOM is satisfied identically. The tensor obtained through the variation of the action corresponds to the obstruction tensor in 6D, which is written in terms of the Riemannian curvature in \cite{1301.7083}. As shown in Eq.\eqref{6DCGaction}, we were able to rewrite it in terms of the Schouten, Cotton and Weyl tensors, such that the vanishing of the obstruction tensor for Einstein manifolds is made manifest.

Six-dimensional CG, described by the LPP action, is also useful to unveil new features of 6D critical gravity at the bi-critical point. As a matter of fact, the decomposition in Einstein and non-Einstein modes presented in Section \ref{5} implies that the Critical Gravity action in Section \ref{6} identically vanishes for any Einstein solution. The fact that Einstein spaces are rendered trivial in this theory needs to be contrasted with thermodynamic properties of generic CG black holes at the bicritical point, along the line of the results derived in the 4D case in \cite{Anastasiou:2017mag} and \cite{Maeda:2018xfn}.

Additionally, in Eq.\eqref{Einstein_TopConf}, the
renormalized Einstein-AdS action in 6D is obtained as a reduced form of a combination of conformal invariants and a topological term. As this prescription linking renormalized Einstein to Conformal gravity is also valid in 4D \cite{1806.10708}, we conjecture that this might be the case in arbitrary even-D. We consider this as a possible criterion to single out the particular combination of conformal invariants that has an Einstein sector, when thought of as a higher-dimensional CG theory. In fact, we believe that the polynomial of the AdS curvature $P_{2n}\left(  \mathcal{F}\right)  $ defined in \cite{1806.10708} is precisely the Einstein-sector of such conformally invariant structure. We shall pursue this idea elsewhere.

\section{Acknowledgments}
It is a pleasure to thank Nicolas Boulanger, Danilo D\'iaz and Stefan Theisen for useful discussions. This work was funded in part by FONDECYT Grants No.~3190314 \emph{Holographic Complexity from anti-de Sitter gravity} (G.A.), No.~3180620 \emph{Entanglement Entropy and AdS gravity} (I.A.) and No.~1170765 \emph{Boundary dynamics in anti-de Sitter gravity and gauge/gravity duality} (R.O.).

\appendix

\section{CG evaluated on Einstein spacetimes}
\label{EinsteinCG}
As shown in Section \ref{2}, both the Cotton tensor \eqref{Cotton} and the $X$ part of the Weyl tensor \eqref{Xcontrib} vanish when evaluated in Einstein spacetimes. Applying these considerations in \eqref{6DCGaction}, we have
\begin{equation}
I_{CG}\left[ E\right]   =\alpha
{\displaystyle\int\limits_{M}}
d^{6}x\sqrt{-\mathcal{G}}\left(  \frac{1}{4!}\delta_{\mu_{1} \ldots \mu_{6}}^{\nu_{1} \ldots \nu_{6}} \mathcal{F}_{\nu_{1} \nu_{2}}^{\mu_{1} \mu_{2}} \mathcal{F}_{\nu_{3} \nu_{4}}^{\mu_{3} \mu_{4}} \mathcal{F}_{\nu_{5} \nu_{6}}^{\mu_{5} \mu_{6}}-\frac{1}{4 \ell^2} \delta_{\mu_{1} \ldots \mu_{5}}^{\nu_{1} \ldots \nu_{5}}\mathcal{F}_{\nu_{1} \nu_{2}}^{\mu_{1} \mu_{2}} \mathcal{F}_{\nu_{3} \nu_{4}}^{\mu_{3} \mu_{4}} \delta_{\nu_{5}}^{\mu_{5}}\right) - \alpha T_{\left(  E\right)  }
\end{equation}
where
\begin{equation}
T_{\left(  E\right)  }= {\displaystyle\int\limits_{\partial M}}
d^{5}x\sqrt{-h}n^{\mu}\left(  \mathcal{F}_{\nu\sigma}^{\kappa\lambda}\nabla_{\mu
} \mathcal{F}_{\kappa\lambda}^{\nu\sigma} \right) .
\end{equation}
Thus, one may rewrite the above expression as
\begin{equation}
I_{CG}\left[ E\right] =-4! \alpha \left(
{\displaystyle\int\limits_{M}}
d^{6}x\sqrt{-\mathcal{G}}P_{6}\left(  \mathcal{F}\right)  +\frac{1}{4!}T_{\left(
E\right)  }\right) , 
\end{equation}
where $P_{6}\left(  \mathcal{F}\right)$  is given in Eq.\eqref{P6}. Here, the polynomial in the AdS curvature tensor, is nothing more than an identity that expresses the topologically renormalized Einstein-Hilbert action, leading to the following expression:
\begin{equation}
I_{CG}\left[ E\right] = \frac{-4!\alpha \left(  16\pi G_{N}\right)}{\ell^4}  \left\{  \frac{1}{16\pi G_{N}}\left[
{\displaystyle\int\limits_{M}}
d^{6}x\sqrt{-\mathcal{G}} \left(  R-2\Lambda  +c_{6}\mathcal{E}
_{6}\right)  \right]  +\frac{\ell^4 T_{\left(  E\right)  }}{4!\left(  16\pi
G_{N}\right)  }\right\} ,
\end{equation}
where $c_{6} = - \frac{\ell^4}{72}$ and $\mathcal{E}_{6}$ is the 6D Euler density, which reads
\begin{equation}
\mathcal{E}_{6} =\frac{1}{2^3}\delta_{\nu_{1}\ldots\nu_{6}}^{\mu
_{1}\ldots\mu_{6}} R_{\mu_{1}\mu_{2}}^{\nu_{1}\nu_{2}} R_{\mu_{3}\mu_{4}}^{\nu_{3}\nu_{4}} R_{\mu_{5}\mu_{6}}^{\nu_{5}\nu_{6}} .
\end{equation}

At this point, the Euler theorem allows us to express $\mathcal{E}_{6}$ in terms of the Chern form $B_5$ as
\begin{equation}
I_{CG} \left[ E\right]= \frac{-384 \pi G_{N}\alpha}{\ell^4}  \left\{  \frac{1}{16\pi G_{N}}\left[
{\displaystyle\int\limits_{M}}
d^{6}x\sqrt{-\mathcal{G}}  \left(  R-2\Lambda  \right)  -\frac{\ell^4}{72}%
{\displaystyle\int\limits_{\partial M}}
B_{5}\right]  +\frac{\ell^4 T_{\left(  E\right)  }}{384 \pi G_{N}
}-\frac{ \pi^2 \ell^4 \chi\left(  M\right)  }{
3 G_{N}}\right\} .
\end{equation}
The expression in the brackets is the definition of the Kounterterm-renormalized Einstein-AdS action, leading to
\begin{equation}
I_{CG}\left[ E\right] =\frac{-384\pi G_{N}\alpha}{\ell^4}  \left\{  \left[  I_{EH}^{KT}+\frac
{\ell^4 T_{\left(  E\right)  }}{384\pi G_{N}}\right]  -\frac
{\pi^2 \ell^4 \chi\left(M\right) }{3 G_{N}%
}\right\} .
\end{equation}

Finally, considering that the total derivative term depending on $T_{\left(E\right)  }$ precisely cancels the extra divergence present in the Kounterterm-renormalized Einstein-AdS action when evaluated on manifolds whose boundary is not conformally flat (see e.g. \cite{Anastasiou:2020zwc}), we have that
\begin{equation}
I_{EH}^{ren}=I_{EH}^{KT}+\frac{\ell^4 T_{\left(  E\right)  }}{384\pi G_{N}},
\label{ModifiedKounter}
\end{equation}
where $I_{EH}^{ren}$ is the fully renormalized action. As a consequence, the Einstein sector of the 6DCG action and the renormalized Einstein-AdS action are related by%
\begin{equation}
I_{EH}^{ren}   =\left(-  \frac{\ell^4}{384 \pi G_{N} \alpha}\right)
I_{CG}\left[ E\right]+\frac{\pi^2 \ell^4}{3G_{N}}\chi\left(M\right).
\end{equation}
Therefore, we remark that the two actions match, up to the Euler characteristic, when $\alpha=\alpha_{E}=-\frac{\ell^4}{384 \pi G_{N}}$.

\section{Divergence cancellation from the 6D CG boundary term}
\label{6DCGdivcancel}

As discussed in \cite{Anastasiou:2020zwc}, the Kounterterms fail to renormalize Einstein-AdS actions on spacetimes that do not have conformally flat boundaries, as the corresponding (partially) renormalized action has remaining divergences. Here we show that the boundary term coming from the conformal invariant action $L_{CG}$, when evaluated on the Einstein sector, precisely cancels the remaining divergence in the $6D$ case.

In particular, we show that
$\frac{\ell^4 T_{\left(  E\right)  }}{384 \pi G_{N}}$ cancels the remaining divergence in $I_{EH}^{KT}$, such that the fully-renormalized EH action can always be written as a combination of a topological and a conformal invariant. To see this, we consider that the $\left\vert \mathcal{W}\left[  h\right] \right\vert ^{2}$ divergence in $I_{EH}^{KT}$ (see Ref.\cite{Anastasiou:2020zwc}) comes from the squared term in the AdS curvature tensor $\mathcal{F}$ in the polynomial $P_{6} \left(\mathcal{F}\right)$, such that%

\begin{align}
I_{div}  &  =\frac{1}{16\pi G_{N}}\frac{\ell^2}{2\left(  4!\right)  }
{\displaystyle\int\limits_{M}}
d^{6}x\sqrt{-\mathcal{G}}\delta_{\mu_{1} \ldots \mu_{4}}^{\nu_{1} \ldots \nu_{4}}\mathcal{F}_{\nu_{1} \nu_{2}}^{\mu_{1}\mu_{2}} \mathcal{F}_{\nu_{1} \nu_{2}}^{\mu_{1}\mu_{2}}-\text{finite}\nonumber\\
&  =\frac{\ell^2}{16\pi G_{N}}\frac{1}{12}
{\displaystyle\int\limits_{M}}
d^{6}x\sqrt{-\mathcal{G}}\left\vert \mathcal{F}\right\vert ^{2}-\text{finite}\nonumber\\
&  =\frac{\ell^3}{16\pi G_{N}}\frac{1}{12}
{\displaystyle\int\limits_{\partial M}}
d^{5}x\sqrt{-h}\left\vert \mathcal{W}\left[  h\right]  \right\vert ^{2}.
\end{align}
Passing from the second to the third line, it is crucial the fact that for Einstein spacetimes $\mathcal{F}_{\nu\sigma}^{\kappa\lambda} =W_{\nu\sigma}^{\kappa\lambda}$, as shown in Section \ref{3}. As a consequence one may rewrite the boundary term $T_{\left(  E\right)  }$ in terms of the bulk Weyl tensor as 
\begin{equation}
T_{\left(  E\right)  }=
{\displaystyle\int\limits_{\partial M}}
d^{5}x\sqrt{-h}n^{\mu}\left(  W_{\nu\sigma}^{\kappa\lambda}\nabla_{\mu
}W_{\kappa\lambda}^{\nu\sigma}\right)  .
\end{equation}
Considering that the normal vector is given by
\begin{align}
n^{\mu}  &  =\left(  -\frac{1}{N\left(  z\right)  },\vec{0}\right),\nonumber\\
N\left(  z\right)   &  =\frac{\ell}{z},
\end{align}
the asymptotic analysis for the boundary term gives
\begin{align}
T_{E}  &  =
{\displaystyle\int\limits_{\partial M}}
d^{5}x\sqrt{-h}n^{\mu}\frac{1}{2}\partial_{\mu}\left\vert W\right\vert
^{2}\nonumber\\
&  =
{\displaystyle\int\limits_{\partial M}}
d^{5}x\sqrt{-h}n^{\mu}\frac{1}{2}\partial_{\mu}\left(  \left\vert
\mathcal{W}\left[  h\right]  \right\vert ^{2}\right) \nonumber\\
&  =- \frac{z}{2 \ell}
{\displaystyle\int\limits_{\partial M}}
d^{5}x\sqrt{-h} \partial_{z}\left(  \frac{z^4}{\ell^4}\left\vert \mathcal{W}
_{0}\right\vert ^{2} +\mathcal{O} \left(z^5\right)\right) \nonumber\\
&  =-\frac{2}{\ell}
{\displaystyle\int\limits_{\partial M}}
d^{5}x\sqrt{-h}\left\vert \mathcal{W}\left[  h\right]  \right\vert ^{2} +\text{finite}.
\end{align}
Therefore, the surface term that appears as a correction to the topologically renormalized Einstein-AdS action \eqref{ModifiedKounter} is expressed as
\begin{equation}
I_{T}=\frac{\ell^4 T_{\left(  E\right)  }}{384 \pi G_{N}  }=-\frac
{\ell^3}{16\pi G_{N}}\frac{1}{12}
{\displaystyle\int\limits_{\partial M}}
d^{5}x\sqrt{-h}\left\vert \mathcal{W}\left[  h\right]  \right\vert
^{2}=-I_{div},
\end{equation}
which implies that it indeed exactly cancels the remaining divergence $I_{div}$ due to the failure of the Kounterterms for 6D asymptotically AdS manifolds without conformally flat boundaries.

\section{6D CG action decomposition (Einstein and non-Einstein)}
\label{CGdecomposition}
Here we provide a detailed analysis of the decomposition of the 6D CG action \eqref{6DCGaction} into an Einstein and a non-Einstein part. This is realized by rewritting the Weyl tensor as \eqref{Weyltensordecomp}. In this case, the first bulk term in the action \eqref{6DCGaction} can be written as
\begin{gather}
L_{CG}^{\left(  1\right)  }=\frac{1}{4!}\delta_{\nu_{1}\ldots\nu_{6}}^{\mu
_{1}\ldots\mu_{6}}W_{\mu_{1}\mu_{2}}^{\nu_{1}\nu_{2}}W_{\mu_{3}\mu_{4}}%
^{\nu_{3}\nu_{4}}W_{\mu_{5}\mu_{6}}^{\nu_{5}\nu_{6}}\nonumber\\
=\frac{1}{4!}\delta_{\nu_{1}\ldots\nu_{6}}^{\mu_{1}\ldots\mu_{6}}\mathcal{F}_{\mu_{1}%
\mu_{2}}^{\nu_{1}\nu_{2}}\mathcal{F}_{\mu_{3}\mu_{4}}^{\nu_{3}\nu_{4}}\mathcal{F}_{\mu_{5}\mu_{6}%
}^{\nu_{5}\nu_{6}}-\frac{1}{8}\delta_{\nu_{1}\ldots\nu_{5}}^{\mu_{1}\ldots
\mu_{5}}\mathcal{F}_{\mu_{1}\mu_{2}}^{\nu_{1}\nu_{2}}\mathcal{F}_{\mu_{3}\mu_{4}}^{\nu_{3}\nu_{4}%
}H_{\mu_{5}}^{\nu_{5}}+\frac{1}{4}\delta_{\nu_{1}\ldots\nu_{4}}^{\mu_{1}%
\ldots\mu_{4}}\left[ \mathcal{F}_{\mu_{1}\mu_{2}}^{\nu_{1}\nu_{2}}H_{\mu_{3}}^{\nu_{3}%
}H_{\mu_{4}}^{\nu_{4}}\right. \nonumber\\
\left.  -2\left(  \frac{1}{\ell^{2}}+\frac{1}{30}R\right)  \mathcal{F}_{\mu_{1}\mu_{2}%
}^{\nu_{1}\nu_{2}}\mathcal{F}_{\mu_{3}\mu_{4}}^{\nu_{3}\nu_{4}}\right]  +\delta_{\nu
_{1}\nu_{2}\nu_{3}}^{\mu_{1}\mu_{2}\mu_{3}}\left[  3\left(  \frac{1}{\ell^{2}%
}+\frac{1}{30}R\right) \mathcal{F}_{\mu_{1}\mu_{2}}^{\nu_{1}\nu_{2}}H_{\mu_{3}}%
^{\nu_{3}}-\frac{1}{4}H_{\mu_{1}}^{\nu_{1}}H_{\mu_{2}}^{\nu_{2}}H_{\mu_{3}%
}^{\nu_{3}}\right] \nonumber\\
-6\left(  \frac{1}{\ell^{2}}+\frac{1}{30}R\right)  \delta_{\nu_{1}\nu_{2}%
}^{\mu_{1}\mu_{2}}H_{\mu_{1}}^{\nu_{1}}H_{\mu_{2}}^{\nu_{2}}+480\left(
\frac{1}{\ell^{2}}+\frac{1}{30}R\right)  ^{3}.
\label{LCG1}%
\end{gather}
For the second bulk term we get
\begin{gather}
L_{CG}^{\left(  2\right)  }=\frac{1}{2}\delta_{\nu_{1}\ldots\nu_{5}}^{\mu
_{1}\ldots\mu_{5}}W_{\mu_{1}\mu_{2}}^{\nu_{1}\nu_{2}}W_{\mu_{3}\mu_{4}}%
^{\nu_{3}\nu_{4}}S_{\mu_{5}}^{\nu_{5}} \nonumber\\
=\frac{1}{8}\delta_{\nu_{1}\ldots\nu_{5}}^{\mu_{1}\ldots\mu_{5}}\mathcal{F}_{\mu_{1}%
\mu_{2}}^{\nu_{1}\nu_{2}}\mathcal{F}_{\mu_{3}\mu_{4}}^{\nu_{3}\nu_{4}}H_{\mu_{5}}%
^{\nu_{5}}+\frac{1}{2}\delta_{\nu_{1}\ldots\nu_{4}}^{\mu_{1}\ldots\mu_{4}%
}\mathcal{F}_{\mu_{1}\mu_{2}}^{\nu_{1}\nu_{2}}\left(  \frac{1}{3}\mathcal{F}_{\mu_{3}\mu_{4}}%
^{\nu_{3}\nu_{4}}S-H_{\mu_{3}}^{\nu_{3}}H_{\mu_{4}}^{\nu_{4}}\right)-24R\left(  \frac{1}{\ell^{2}}+\frac{1}{30}R\right)^{2}
\nonumber\\
+\delta_{\nu_{1}\nu_{2}\nu_{3}}^{\mu_{1}\mu_{2}\mu_{3}}\left[  \frac{3}%
{4}H_{\mu_{1}}^{\nu_{1}}H_{\mu_{2}}^{\nu_{2}}-\left(  \frac{3}{\ell^{2}}%
+\frac{1}{5}R\right) \mathcal{F}_{\mu_{1}\mu_{2}}^{\nu_{1}\nu_{2}}\right]  H_{\mu_{3}%
}^{\nu_{3}}+3\left(  \frac{4}{\ell^{2}}+\frac{1}{5}R\right)  \delta_{\nu
_{1}\nu_{2}}^{\mu_{1}\mu_{2}}H_{\mu_{1}}^{\nu_{1}}H_{\mu_{2}}^{\nu_{2}%
}.
\label{LCG2}
\end{gather}
Finally, the last bulk term acquires the form
\begin{gather}
L_{CG}^{\left(  3\right)  }=8C_{\nu
\lambda}^{\mu}C_{\mu}^{\nu\lambda}=4\delta_{\nu_{1}\nu_{2}\nu_{3}}^{\mu_{1}\mu_{2}\mu_{3}}C_{\mu_{1}}^{\nu_{1}\nu_{2}}C_{\mu_{2}\mu_{3}}^{\nu_{3}
}\nonumber\\
=\nabla_{\lambda}H_{\nu}^{\mu}\nabla^{\lambda
}H_{\mu}^{\nu}-\nabla_{\lambda}H_{\nu}^{\mu}\nabla^{\nu}H_{\mu}^{\lambda
}-\frac{1}{5}\nabla_{\mu}H_{\nu}^{\mu}\nabla^{\lambda}H_{\lambda}^{\nu}.
\label{LCG3}
\end{gather}
In total, summing up all the bulk contributions considered in \eqref{LCG1}, \eqref{LCG2} and \eqref{LCG3} leads to the following expression:
\begin{gather}
L_{CG}^{\text{bulk}}=L_{CG}^{\left(  1\right)  }+L_{CG}^{\left(  2\right)  }
+L_{CG}^{\left(  3\right)  }\nonumber\\
=\frac{1}{4!}\delta_{\nu_{1}\ldots\nu_{6}}^{\mu_{1}\ldots\mu_{6}}\mathcal{F}_{\mu_{1}%
\mu_{2}}^{\nu_{1}\nu_{2}}\mathcal{F}_{\mu_{3}\mu_{4}}^{\nu_{3}\nu_{4}}\mathcal{F}_{\mu_{5}\mu_{6}%
}^{\nu_{5}\nu_{6}}-\frac{1}{2\ell^{2}}\delta_{\nu_{1}\ldots\nu_{4}}^{\mu
_{1}\ldots\mu_{4}}\mathcal{F}_{\mu_{1}\mu_{2}}^{\nu_{1}\nu_{2}}\mathcal{F}_{\mu_{3}\mu_{4}}
^{\nu_{3}\nu_{4}}\nonumber\\
-\frac{1}{4}\delta_{\nu_{1}\ldots\nu_{4}}^{\mu_{1}\ldots\mu_{4}}\mathcal{F}_{\mu_{1}
\mu_{2}}^{\nu_{1}\nu_{2}}H_{\mu_{3}}^{\nu_{3}}H_{\mu_{4}}^{\nu_{4}}+\frac
{1}{2}\delta_{\nu_{1}\nu_{2}\nu_{3}}^{\mu_{1}\mu_{2}\mu_{3}}\left(  H_{\mu
_{1}}^{\nu_{1}}H_{\mu_{2}}^{\nu_{2}}-\frac{1}{5}\mathcal{F}_{\mu_{1}\mu_{2}}^{\nu_{1}
\nu_{2}}\right)  H_{\mu_{3}}^{\nu_{3}}\nonumber\\
+2\left(  \frac{3}{\ell^{2}}+\frac{1}{5}R\right)  \delta_{\nu_{1}\nu_{2}}
^{\mu_{1}\mu_{2}}H_{\mu_{1}}^{\nu_{1}}H_{\mu_{2}}^{\nu_{2}}+\nabla_{\lambda
}H_{\nu}^{\mu}\nabla^{\lambda}H_{\mu}^{\nu}-\nabla_{\lambda}H_{\nu}^{\mu
}\nabla^{\nu}H_{\mu}^{\lambda}-\frac{1}{5}\nabla_{\mu}H_{\nu}^{\mu}
\nabla^{\lambda}H_{\lambda}^{\nu}\nonumber\\
-8\left(  R-\frac{60}{\ell^{2}}\right)  \left(  \frac{1}{\ell^{2}}+\frac
{1}{30}R\right)^{2}.
\label{LCGbulk}
\end{gather}
In order to conclude the off-shell decomposition of the LPP action in terms of the Einstein and non-Einstein modes, we have to analyze the boundary term of \eqref{6DCGaction}. Thus, one writes
\begin{gather}
L_{CG}^{\text{boundary}}=8W_{\mu}^{\ \kappa\lambda\nu}C_{\kappa\lambda\nu}%
-W_{\nu\sigma}^{\kappa\lambda}\nabla_{\mu}W_{\kappa\lambda}^{\nu\sigma} \nonumber\\
=4\mathcal{F}_{\mu\kappa}^{\lambda\nu}\nabla_{\nu}H_{\lambda}^{\kappa}-H_{\mu}^{\lambda
}\nabla_{\nu}H_{\lambda}^{\nu}+H_{\kappa}^{\lambda}\nabla_{\mu}H_{\lambda
}^{\kappa}-H_{\kappa}^{\nu}\nabla_{\nu}H_{\mu}^{\kappa}-\frac{1}{4}\delta
_{\nu_{1}\ldots\nu_{4}}^{\mu_{1}\ldots\mu_{4}}\mathcal{F}_{\mu_{1}\mu_{2}}^{\nu_{1}%
\nu_{2}}\nabla_{\mu}\mathcal{F}_{\mu_{3}\mu_{4}}^{\nu_{3}\nu_{4}}\nonumber\\
+\frac{3}{4}\delta_{\nu_{1}\nu_{2}\nu_{3}}^{\mu_{1}\mu_{2}\mu_{3}}\left(
\mathcal{F}_{\mu_{1}\mu_{2}}^{\nu_{1}\nu_{2}}\nabla_{\mu}H_{\mu_{3}}^{\nu_{3}}%
+H_{\mu_{1}}^{\nu_{1}}\nabla_{\mu}\mathcal{F}_{\mu_{2}\mu_{3}}^{\nu_{2}\nu_{3}}\right)
-3\delta_{\nu_{1}\nu_{2}}^{\mu_{1}\mu_{2}}H_{\mu_{1}}^{\nu_{1}}\nabla_{\mu
}H_{\mu_{2}}^{\nu_{2}}+\frac{16}{15}\left(  \frac{30}{\ell^{2}}+R\right)
\nabla_{\nu}H_{\mu}^{\nu}.
\label{LCGbound}
\end{gather}
Summing the bulk and the boundary contributions, Eqs. \eqref{LCGbulk} and \eqref{LCGbound} respectively, the LPP action obtains the form
\begin{gather}
I_{CG}=\alpha {\displaystyle\int\limits_{ M}}d^{6}x\sqrt{-\mathcal{G}}\left[  \frac{1}{4!}\delta_{\nu_{1}%
\ldots\nu_{6}}^{\mu_{1}\ldots\mu_{6}}\mathcal{F}_{\mu_{1}\mu_{2}}^{\nu_{1}\nu_{2}}%
\mathcal{F}_{\mu_{3}\mu_{4}}^{\nu_{3}\nu_{4}}\mathcal{F}_{\mu_{5}\mu_{6}}^{\nu_{5}\nu_{6}}%
-\frac{1}{2\ell^{2}}\delta_{\nu_{1}\ldots\nu_{4}}^{\mu_{1}\ldots\mu_{4}}%
\mathcal{F}_{\mu_{1}\mu_{2}}^{\nu_{1}\nu_{2}}\mathcal{F}_{\mu_{3}\mu_{4}}^{\nu_{3}\nu_{4}}\right.
\nonumber\\
-\frac{1}{4}\delta_{\nu_{1}\ldots\nu_{4}}^{\mu_{1}\ldots\mu_{4}}\mathcal{F}_{\mu_{1}%
\mu_{2}}^{\nu_{1}\nu_{2}}H_{\mu_{3}}^{\nu_{3}}H_{\mu_{4}}^{\nu_{4}}+\frac
{1}{2}\delta_{\nu_{1}\nu_{2}\nu_{3}}^{\mu_{1}\mu_{2}\mu_{3}}\left(  H_{\mu
_{1}}^{\nu_{1}}H_{\mu_{2}}^{\nu_{2}}-\frac{1}{5}\mathcal{F}_{\mu_{1}\mu_{2}}^{\nu_{1}%
\nu_{2}}\right)  H_{\mu_{3}}^{\nu_{3}}\nonumber\\
+\frac{2}{5}\left(  \frac{15}{\ell^{2}}+R\right)  \delta_{\nu_{1}\nu_{2}}%
^{\mu_{1}\mu_{2}}H_{\mu_{1}}^{\nu_{1}}H_{\mu_{2}}^{\nu_{2}}+\nabla_{\lambda
}H_{\nu}^{\mu}\nabla^{\lambda}H_{\mu}^{\nu}-\nabla_{\lambda}H_{\nu}^{\mu
}\nabla^{\nu}H_{\mu}^{\lambda}\nonumber\\
\left.  -\frac{1}{5}\nabla_{\mu}H_{\nu}^{\mu}\nabla^{\lambda}H_{\lambda}^{\nu
}-\frac{2}{225}\left(  R-\frac{60}{\ell^{2}}\right)  \left(  \frac{30}%
{\ell^{2}}+R\right)^{2}\right] \nonumber\\
+\alpha {\displaystyle\int\limits_{\partial M}}d^{5}x\sqrt{-h}n^{\mu}\left[  -\frac{1}{4}\delta
_{\nu_{1}\ldots\nu_{4}}^{\mu_{1}\ldots\mu_{4}}\mathcal{F}_{\mu_{1}\mu_{2}}^{\nu_{1}%
\nu_{2}}\nabla_{\mu}\mathcal{F}_{\mu_{3}\mu_{4}}^{\nu_{3}\nu_{4}}+\frac{3}{4}\delta
_{\nu_{1}\nu_{2}\nu_{3}}^{\mu_{1}\mu_{2}\mu_{3}}\left( \mathcal{F}_{\mu_{1}\mu_{2}%
}^{\nu_{1}\nu_{2}}\nabla_{\mu}H_{\mu_{3}}^{\nu_{3}}+H_{\mu_{1}}^{\nu_{1}%
}\nabla_{\mu}\mathcal{F}_{\mu_{2}\mu_{3}}^{\nu_{2}\nu_{3}}\right)  \right. \nonumber\\
\left.  +4\mathcal{F}_{\mu\kappa}^{\lambda\nu}\nabla_{\nu}H_{\lambda}^{\kappa}-H_{\mu
}^{\lambda}\nabla_{\nu}H_{\lambda}^{\nu}-H_{\kappa}^{\nu}\nabla_{\nu}H_{\mu
}^{\kappa}-2H_{\kappa}^{\lambda}\nabla_{\mu}H_{\lambda}^{\kappa}+\frac{16}%
{15}\left(  \frac{30}{\ell^{2}}+R\right)  \nabla_{\nu}H_{\mu}^{\nu}\right]  .
\label{6DCGdecomp}%
\end{gather}
Here $\alpha$ is the overall factor coming from the relation between the coefficients of the conformal invariants that allow the LPP action to admit Einstein spacetimes as solutions.

\section{Obstruction tensor derivation in 6D CG}
\label{EOM}
For the sake of simplicity we rewrite the bulk Lagrangian in \eqref{6DCGaction} as
\begin{gather}
\frac{1}{4 !}\delta _{\nu _{1}\ldots \nu _{6}}^{\mu _{1}\ldots \mu _{6}} W_{\mu _{1}\mu _{2}}^{\nu _{1}\nu _{2}}W_{\mu _{3}\mu _{4}}^{\nu _{3}\nu _{4}}W_{\mu _{5}\mu _{6}}^{\nu _{5}\nu _{6}} +\frac{1}{2}\delta _{\nu _{1}\ldots \nu _{5}}^{\mu _{1}\ldots \mu _{5}} W_{\mu _{1}\mu _{2}}^{\nu _{1}\nu _{2}}W_{\mu _{3}\mu _{4}}^{\nu _{3}\nu _{4}}S_{\mu _{5}}^{\nu _{5}} +8C^{\mu \nu \lambda }C_{\mu \nu \lambda } = \nonumber  \\
-\frac{1}{12}\delta _{\nu _{1}\ldots \nu _{6}}^{\mu _{1}\ldots \mu _{6}}W_{\mu _{1}\mu _{2}}^{\nu _{1}\nu _{2}}W_{\mu _{3}\mu _{4}}^{\nu _{3}\nu _{4}}W_{\mu _{5}\mu _{6}}^{\nu _{5}\nu _{6}} +\frac{1}{8}\delta _{\nu _{1}\ldots \nu _{6}}^{\mu _{1}\ldots \mu _{6}}W_{\mu _{1}\mu _{2}}^{\nu _{1}\nu _{2}}W_{\mu _{3}\mu _{4}}^{\nu _{3}\nu _{4}}R_{\mu _{5}\mu _{6}}^{\nu _{5}\nu _{6}} +8C^{\mu \nu \lambda }C_{\mu \nu \lambda } .
\end{gather}
Since our focus is the EOM of the theory, we drop the surface terms of the action. Thus, the variation of the action now reads
\begin{gather}
\delta I_{CG}^{\text{bulk}}=\int\limits_{M}d^{6}x\sqrt{-\mathcal{G}}\bigg\{\frac{1}%
{2}\left[  \frac{-1}{12}\delta_{\nu_{1}\ldots\nu_{6}}^{\mu_{1}\ldots\mu_{6}%
}W_{\mu_{1}\mu_{2}}^{\nu_{1}\nu_{2}}W_{\mu_{3}\mu_{4}}^{\nu_{3}\nu_{4}}%
W_{\mu_{5}\mu_{6}}^{\nu_{5}\nu_{6}}+\frac{1}{8}\delta_{\nu_{1}\ldots\nu_{6}%
}^{\mu_{1}\ldots\mu_{6}}W_{\mu_{1}\mu_{2}}^{\nu_{1}\nu_{2}}W_{\mu_{3}\mu_{4}%
}^{\nu_{3}\nu_{4}}R_{\mu_{5}\mu_{6}}^{\nu_{5}\nu_{6}}+8C^{\mu\nu\lambda}%
C_{\mu\nu\lambda}\right]  \left(  \mathcal{G}^{-1}\delta\mathcal{G}\right)
\nonumber\\
-\frac{1}{12}\delta_{\nu_{1}\ldots\nu_{6}}^{\mu_{1}\ldots\mu_{6}}\delta\left(
W_{\mu_{1}\mu_{2}}^{\nu_{1}\nu_{2}}W_{\mu_{3}\mu_{4}}^{\nu_{3}\nu_{4}}%
W_{\mu_{5}\mu_{6}}^{\nu_{5}\nu_{6}}\right)  +\frac{1}{8}\delta_{\nu_{1}%
\ldots\nu_{6}}^{\mu_{1}\ldots\mu_{6}}\delta\left(  W_{\mu_{1}\mu_{2}}^{\nu
_{1}\nu_{2}}W_{\mu_{3}\mu_{4}}^{\nu_{3}\nu_{4}}R_{\mu_{5}\mu_{6}}^{\nu_{5}%
\nu_{6}}\right)  +8\delta\left(  C^{\mu\nu\lambda}C_{\mu\nu\lambda}\right)
\bigg\},\label{varcgbulk}%
\end{gather}
For the explicit derivation of the EOM, the following variations are going to be used:
\begin{gather}
\delta  W_{\nu _{1} \nu _{2}}^{\mu _{1} \mu _{2}}  = \delta  R_{\nu _{1} \nu _{2}}^{\mu _{1} \mu _{2}} -\left (\delta _{\nu _{1}}^{\mu _{1}}\delta S_{\nu _{2}}^{\mu _{2}} -\delta _{\nu _{1}}^{\mu _{2}}\delta S_{\nu _{2}}^{\mu _{1}} -\delta _{\nu _{2}}^{\mu _{1}}\delta S_{\nu _{1}}^{\mu _{2}} +\delta _{\nu _{2}}^{\mu _{2}}\delta S_{\nu _{1}}^{\mu _{1}}\right ) , \nonumber \\
\delta    R_{\nu _{1} \nu _{2}}^{\mu _{1} \mu _{2}} =\mathcal{G}^{\mu _{2}\kappa }\delta R_{\kappa \nu _{1}\nu _{2}}^{\mu _{1}} +R_{\kappa \nu _{1}\nu _{2}}^{\mu _{1}}\delta \mathcal{G}^{\mu _{2}\kappa } , \nonumber \\
\delta R_{\ \kappa \nu _{1}\nu _{2}}^{\mu _{1}} = \nabla _{\nu _{1}}\delta \Gamma _{\kappa  \nu _{2}}^{\mu _{1}} - \nabla _{\nu _{2}}\delta \Gamma _{\kappa  \nu _{1}}^{\mu _{1}} , \nonumber \\
\delta \Gamma _{\kappa  \nu _{2}}^{\mu _{1}} =\frac{1}{2}\mathcal{G}^{\mu _{1} \sigma }\left ( \nabla _{\kappa }\delta  \mathcal{G}_{\sigma  \nu _{2}} + \nabla _{\nu _{2}}\delta  \mathcal{G}_{\kappa  \sigma } - \nabla _{\sigma }\delta  \mathcal{G}_{\kappa  \nu _{2}}\right ) , \nonumber \\
\delta S_{\nu _{1}}^{\mu _{1}} =\mathcal{G}^{\mu _{1}\kappa }\delta S_{\kappa \nu _{1}} +S_{\kappa \nu _{1}}\delta \mathcal{G}^{\mu _{1}\kappa } , \nonumber \\
\delta S_{\kappa \nu _{1}} =\frac{1}{4}\left (\delta R_{\kappa \nu _{1}} -\frac{1}{10}G_{\kappa \nu _{1}}\delta R -\frac{1}{10}R\delta G_{\kappa \nu _{1}}\right ) , \nonumber \\
\delta R_{\kappa \nu _{1}} = \nabla _{\sigma }\delta \Gamma _{\kappa \nu _{1}}^{\sigma } - \nabla _{\nu _{1}}\delta \Gamma _{\kappa \sigma }^{\sigma } .
\end{gather}

We separate \eqref{varcgbulk} into three pieces. For the first contribution, we write
\begin{gather}
\delta \mathcal{I}_{1} = -\frac{1}{12} \delta _{\nu _{1}\ldots \nu _{6}}^{\mu _{1}\ldots \mu _{6}} \left [\frac{1}{2}W_{\mu _{1}\mu _{2}}^{\nu _{1}\nu _{2}}W_{\mu _{3}\mu _{4}}^{\nu _{3}\nu _{4}}W_{\mu _{5}\mu _{6}}^{\nu _{5}\nu _{6}}\left (\mathcal{G}^{ -1}\delta \mathcal{G}\right ) +3W_{\mu _{1}\mu _{2}}^{\nu _{1}\nu _{2}}W_{\mu _{3}\mu _{4}}^{\nu _{3}\nu _{4}}\delta W_{\mu _{5}\mu _{6}}^{\nu _{5}\nu _{6}}\right ] \nonumber  \\
=\frac{1}{4}\delta _{\nu _{1}\ldots \nu _{6}}^{\mu _{1}\ldots \mu _{6}} W_{\mu _{1}\mu _{2}}^{\nu _{1}\nu _{2}}W_{\mu _{3}\mu _{4}}^{\nu _{3}\nu _{4}}\mathcal{G}^{\kappa \nu _{5}}\delta W_{\ \kappa \mu _{5}\mu _{6}}^{\nu _{6}} .
\label{i1}
\end{gather}

For the second contribution, it follows
\begin{gather}
\delta \mathcal{I}_{2} =\frac{1}{8}\delta _{\nu _{1}\ldots \nu _{6}}^{\mu _{1}\ldots \mu _{6}}W_{\mu _{1}\mu _{2}}^{\nu _{1}\nu _{2}}\left [\frac{1}{2}W_{\mu _{3}\mu _{4}}^{\nu _{3}\nu _{4}}R_{\mu _{5}\mu _{6}}^{\nu _{5}\nu _{6}}\left (\mathcal{G}^{ -1}\delta \mathcal{G} \right ) +2R_{\mu _{3}\mu _{4}}^{\nu _{3}\nu _{4}}\delta W_{\mu _{5}\mu _{6}}^{\nu _{5}\nu _{6}} +W_{\mu _{3}\mu _{4}}^{\nu _{3}\nu _{4}}\delta R_{\mu _{5}\mu _{6}}^{\nu _{5}\nu _{6}}\right ] \nonumber  \\
=\frac{1}{8}\delta _{\nu _{1}\ldots \nu _{6}}^{\mu _{1}\ldots \mu _{6}} W_{\mu _{1}\mu _{2}}^{\nu _{1}\nu _{2}}\left (2R_{\mu _{3}\mu _{4}}^{\nu _{3}\nu _{4}}\delta W_{\ \kappa \mu _{5}\mu _{6}}^{\nu _{5}} +W_{\mu _{3}\mu _{4}}^{\nu _{3}\nu _{4}}\delta R_{\ \kappa \mu _{5}\mu _{6}}^{\nu _{5}}\right )\mathcal{G}^{\kappa \nu _{6}} . 
\label{i2}
\end{gather}
Substituting \eqref{i1} and \eqref{i2} into \eqref{varcgbulk}, we manage to simplify the latter which now reads
\begin{gather}
\delta I_{CG}^{\text{bulk}} =\int\limits_{M}d^{6}x\sqrt{ -\mathcal{G}}\bigg[ -\frac{1}{8}\delta _{\nu _{1}\ldots \nu _{6}}^{\mu _{1}\ldots \mu _{6}} W_{\mu _{1}\mu _{2}}^{\nu _{1}\nu _{2}}W_{\mu _{3}\mu _{4}}^{\nu _{3}\nu _{4}}\mathcal{G}^{\kappa \nu _{5}}\delta R_{\ \kappa \mu _{5}\mu _{6}}^{\nu _{6}} -\delta _{\nu _{1}\ldots \nu _{5}}^{\mu _{1}\ldots \mu _{5}} W_{\mu _{1}\mu _{2}}^{\nu _{1}\nu _{2}}S_{\mu _{3}}^{\nu _{3}}\mathcal{G}^{\kappa \nu _{4}}\delta W_{\ \kappa \mu _{4}\mu _{5}}^{\nu _{5}} \nonumber  \\
+2\delta _{\nu _{1}\nu _{2}\nu _{3}}^{\mu _{1}\mu _{2}\mu _{3}}C_{\mu _{1}}^{\nu _{1}\nu _{2}}C_{\mu _{2}\mu _{3}}^{\nu _{3}}\left (\mathcal{G}^{ -1}\delta \mathcal{G}\right ) +4\delta _{\nu _{1}\nu _{2}\nu _{3}}^{\mu _{1}\mu _{2}\mu _{3}}\delta \left (C_{\mu _{1}}^{\nu _{1}\nu _{2}}C_{\mu _{2}\mu _{3}}^{\nu _{3}}\right )\bigg] .
\label{deltaIbulk}
\end{gather}
Here, the first term of \eqref{deltaIbulk} reads
\begin{gather}
-\frac{1}{8}\delta _{\nu _{1}\ldots \nu _{6}}^{\mu _{1}\ldots \mu _{6}} W_{\mu _{1}\mu _{2}}^{\nu _{1}\nu _{2}}W_{\mu _{3}\mu _{4}}^{\nu _{3}\nu _{4}}\mathcal{G}^{\kappa \nu _{5}}\delta R_{\kappa \mu_{5}\mu_{6}}^{\nu _{6}} = -\delta _{\nu _{1}\ldots \nu _{5}}^{\mu _{1}\ldots \mu _{5}} W_{\mu _{1}\mu _{2}}^{\nu _{1}\nu _{2}}\left ( \nabla ^{\nu _{3}}C_{\mu _{3}\mu _{4}}^{\nu _{4}}\right )\left (\mathcal{G}^{ -1}\delta \mathcal{G}\right )_{\mu _{5}}^{\nu _{5}} +4\delta _{\nu _{1}\ldots \nu _{4}}^{\mu _{1}\ldots \mu _{4}}C_{\mu _{1}}^{\nu _{1}\nu _{2}}C_{\mu _{2}\mu _{3}}^{\nu _{3}}\left (\mathcal{G}^{ -1}\delta \mathcal{G}\right )_{\mu _{4}}^{\nu _{4}} \nonumber  \\
-\frac{1}{4}\delta _{\nu _{1}\ldots \nu _{6}}^{\mu _{1}\ldots \mu _{6}} \nabla _{\mu _{5}}\left (W_{\mu _{1}\mu _{2}}^{\nu _{1}\nu _{2}}W_{\mu _{3}\mu _{4}}^{\nu _{3}\nu _{4}}\mathcal{G}^{\kappa \nu _{5}}\delta \Gamma _{\kappa \mu _{6}}^{\nu _{6}}\right ) -\delta _{\nu _{1}\ldots \nu _{5}}^{\mu _{1}\ldots \mu _{5}} \nabla ^{\nu _{4}}\left [W_{\mu _{1}\mu _{2}}^{\nu _{1}\nu _{2}}C_{\mu _{3}\mu _{4}}^{\nu _{3}}\left (\mathcal{G}^{ -1}\delta \mathcal{G}\right )_{\mu _{5}}^{\nu _{5}}\right ] ,
\end{gather}
where only the first two terms in the RHS contribute to the EOM and the rest are just boundary terms.

As for the second contribution in \eqref{deltaIbulk}, we get
\begin{gather}
-\delta _{\nu _{1}\ldots \nu _{5}}^{\mu _{1}\ldots \mu _{5}} W_{\mu _{1}\mu _{2}}^{\nu _{1}\nu _{2}}S_{\mu _{3}}^{\nu _{3}}\mathcal{G}^{\kappa \nu _{4}}\delta W_{\ \kappa \mu _{4}\mu _{5}}^{\nu _{5}} = -2\delta _{\nu _{1}\ldots \nu _{5}}^{\mu _{1}\ldots \mu _{5}}W_{\mu _{1}\mu _{2}}^{\nu _{1}\nu _{2}}S_{\mu _{3}}^{\nu _{3}}S_{\mu _{4}}^{\nu _{4}}\left (\mathcal{G}^{ -1}\delta \mathcal{G}\right )_{\mu _{5}}^{\nu _{5}} \nonumber  \\
-8\delta _{\nu _{1}\ldots \nu _{4}}^{\mu _{1}\ldots \mu _{4}}\left [\frac{1}{2}C_{\mu _{1}}^{\nu _{1}\nu _{2}}C_{\mu _{2}\mu _{3}}^{\nu _{3}} +S_{\mu _{1}}^{\nu _{1}}\left ( \nabla ^{\nu _{2}}C_{\mu _{2}\mu _{3}}^{\nu _{3}}\right )\right ]\left (\mathcal{G}^{ -1}\delta \mathcal{G}\right )_{\mu _{4}}^{\nu _{4}} -4\delta _{\nu _{1}\ldots \nu _{4}}^{\mu _{1}\ldots \mu _{4}}C_{\mu _{1}}^{\nu _{1}\nu _{2}}C_{\mu _{2}\mu _{3}}^{\nu _{3}}\left (\mathcal{G}^{ -1}\delta \mathcal{G}\right )_{\mu _{4}}^{\nu _{4}} \nonumber  \\
+\delta _{\nu _{1}\ldots \nu _{5}}^{\mu _{1}\ldots \mu _{5}}W_{\mu _{1}\mu _{2}}^{\nu _{1}\nu _{2}}\left ( \nabla ^{\nu _{3}}C_{\mu _{3}\mu _{4}}^{\nu _{4}}\right )\left (\mathcal{G}^{ -1}\delta \mathcal{G}\right )_{\mu _{5}}^{\nu _{5}} +2\delta _{\nu _{1}\ldots \nu _{4}}^{\mu _{1}\ldots \mu _{4}}W_{\mu _{1}\mu _{2}}^{\nu _{1}\nu _{2}}S_{\mu _{3}}^{\nu _{3}}\left [2S_{\mu _{4}}^{\kappa }\left (\mathcal{G}^{ -1}\delta \mathcal{G}\right )_{\kappa }^{\nu _{4}} +S\left (\mathcal{G}^{ -1}\delta \mathcal{G}\right )_{\mu _{4}}^{\nu _{4}}\right ] \nonumber  \\
-\delta _{\nu _{1}\ldots \nu _{4}}^{\mu _{1}\ldots \mu _{4}}\bigg\{\left [\left ( \nabla ^{\nu _{4}} \nabla _{\sigma }W_{\mu _{1}\mu _{2}}^{\nu _{1}\nu _{2}}\right )S_{\mu _{3}}^{\nu _{3}} +\frac{1}{2}\left ( \nabla _{\sigma }W_{\mu _{1}\mu _{2}}^{\nu _{1}\nu _{2}}\right )C_{\mu _{3}}^{\nu _{3}\nu _{4}} +W_{\mu _{1}\mu _{2}}^{\nu _{1}\nu _{2}}\left ( \nabla ^{\nu _{4}} \nabla _{\sigma }S_{\mu _{3}}^{\nu _{3}}\right )\right ]\left (\mathcal{G}^{ -1}\delta \mathcal{G}\right )_{\mu _{4}}^{\sigma } \nonumber  \\
+\left [( \nabla _{\mu _{4}} \nabla ^{\sigma }W_{\mu _{1}\mu _{2}}^{\nu _{1}\nu _{2}})S_{\mu _{3}}^{\nu _{3}} +\frac{1}{2}( \nabla ^{\sigma }W_{\mu _{1}\mu _{2}}^{\nu _{1}\nu _{2}})C_{\mu _{3}\mu _{4}}^{\nu _{3}} +W_{\mu _{1}\mu _{2}}^{\nu _{1}\nu _{2}}\left ( \nabla _{\mu _{4}} \nabla ^{\sigma }S_{\mu _{3}}^{\nu _{3}}\right )\right ]\left (\mathcal{G}^{ -1}\delta \mathcal{G}\right )_{\sigma }^{\nu _{4}} \nonumber  \\
-\Big [\left ( \nabla ^{\sigma } \nabla _{\sigma }W_{\mu _{1}\mu _{2}}^{\nu _{1}\nu _{2}}\right ) +2\left ( \nabla ^{\sigma }W_{\mu _{1}\mu _{2}}^{\nu _{1}\nu _{2}}\right )\left ( \nabla _{\sigma }S_{\mu _{3}}^{\nu _{3}}\right ) +W_{\mu _{1}\mu _{2}}^{\nu _{1}\nu _{2}}\left ( \nabla ^{\sigma } \nabla _{\sigma }S_{\mu _{3}}^{\nu _{3}}\right )\Big]\left (\mathcal{G}^{ -1}\delta \mathcal{G}\right )_{\mu _{4}}^{\nu _{4}} \nonumber \\
+\frac{1}{2}W_{\mu _{1}\mu _{2}}^{\nu _{1}\nu _{2}}\left ( \nabla ^{\nu _{3}}C_{\mu _{3}\mu _{4}}^{\nu _{4}}\right )\left (\mathcal{G}^{ -1}\delta \mathcal{G}\right )\bigg\} \nonumber  \\
+6\delta _{\nu _{1}\nu _{2}\nu _{3}}^{\mu _{1}\mu _{2}\mu _{3}}\left \{\Big [C_{\mu _{1}}^{\nu _{1}\nu _{2}}C_{\mu _{2}\mu _{3}}^{\nu _{3}} +S_{\mu _{1}}^{\nu _{1}}\left ( \nabla ^{\nu _{2}}C_{\mu _{2}\mu _{3}}^{\nu _{3}}\right )\Big ]\left (\mathcal{G}^{ -1}\delta \mathcal{G}\right ) +\left ( \nabla _{\sigma }S_{\mu _{1}}^{\nu _{1}}\right )C_{\mu _{2}}^{\nu _{2}\nu _{3}}\left (\mathcal{G}^{ -1}\delta \mathcal{G}\right )_{\mu _{3}}^{\sigma } \right. \nonumber  \\
\left. +\left ( \nabla ^{\sigma }S_{\mu _{1}}^{\nu _{1}}\right )C_{\mu _{2}\mu _{3}}^{\nu _{2}}\left (\mathcal{G}^{ -1}\delta \mathcal{G}\right )_{\sigma }^{\nu _{3}}\right \} 
-2\delta _{\nu _{1}\ldots \nu _{5}}^{\mu _{1}\ldots \mu _{5}} \nabla _{\mu _{4}}\left [W_{\mu _{1}\mu _{2}}^{\nu _{1}\nu _{2}}S_{\mu _{3}}^{\nu _{3}}\mathcal{G}^{\kappa \nu _{4}}\delta \Gamma _{\kappa \mu _{5}}^{\nu _{5}}\right ] +\delta _{\nu _{1}\ldots \nu _{5}}^{\mu _{1}\ldots \mu _{5}} \nabla ^{\nu _{4}}\left [W_{\mu _{1}\mu _{2}}^{\nu _{1}\nu _{2}}C_{\mu _{3}\mu _{4}}^{\nu _{3}}\left (\mathcal{G}^{ -1}\delta \mathcal{G}\right )_{\mu _{5}}^{\nu _{5}}\right ] \nonumber  \\
-8\delta _{\nu _{1}\ldots \nu _{4}}^{\mu _{1}\ldots \mu _{4}} \nabla ^{\nu _{3}}\left [S_{\mu _{1}}^{\nu _{1}}C_{\mu _{2}\mu _{3}}^{\nu _{2}}\left (\mathcal{G}^{ -1}\delta \mathcal{G}\right )_{\mu _{4}}^{\nu _{4}}\right ] +6\delta _{\nu _{1}\nu _{2}\nu _{3}}^{\mu _{1}\mu _{2}\mu _{3}} \nabla ^{\nu _{3}}\left [S_{\mu _{1}}^{\nu _{1}}C_{\mu _{2}\mu _{3}}^{\nu _{2}}\left (\mathcal{G}^{ -1}\delta \mathcal{G}\right )\right ] \nonumber  \\
+\delta _{\nu _{1}\ldots \nu _{4}}^{\mu _{1}\ldots \mu _{4}}\bigg\{ \nabla _{\mu _{4}}\left \{\Big [( \nabla ^{\sigma }W_{\mu _{1}\mu _{2}}^{\nu _{1}\nu _{2}})S_{\mu _{3}}^{\nu _{3}} +W_{\mu _{1}\mu _{2}}^{\nu _{1}\nu _{2}}\left ( \nabla ^{\sigma }S_{\mu _{3}}^{\nu _{3}}\right )\Big ]\left (\mathcal{G}^{ -1}\delta \mathcal{G}\right )_{\sigma }^{\nu _{4}} +2W_{\mu _{1}\mu _{2}}^{\nu _{1}\nu _{2}}S_{\mu _{3}}^{\nu _{3}}\mathcal{G}^{\kappa \nu _{4}}\delta \Gamma _{\kappa \sigma }^{\sigma }\right \} \nonumber  \\
- \nabla _{\sigma }\left \{\Big [( \nabla ^{\sigma }W_{\mu _{1}\mu _{2}}^{\nu _{1}\nu _{2}})S_{\mu _{3}}^{\nu _{3}} +W_{\mu _{1}\mu _{2}}^{\nu _{1}\nu _{2}}\left ( \nabla ^{\sigma }S_{\mu _{3}}^{\nu _{3}}\right )\Big ]\left (\mathcal{G}^{ -1}\delta \mathcal{G}\right )_{\mu _{4}}^{\nu _{4}} +2W_{\mu _{1}\mu _{2}}^{\nu _{1}\nu _{2}}S_{\mu _{3}}^{\nu _{3}}\mathcal{G}^{\kappa \nu _{4}}\delta \Gamma _{\kappa \mu _{4}}^{\sigma }\right \} \nonumber  \\
+ \nabla ^{\nu _{4}}\left \{\Big [( \nabla _{\sigma }W_{\mu _{1}\mu _{2}}^{\nu _{1}\nu _{2}})S_{\mu _{3}}^{\nu _{3}} +W_{\mu _{1}\mu _{2}}^{\nu _{1}\nu _{2}}\left ( \nabla _{\sigma }S_{\mu _{3}}^{\nu _{3}}\right )\Big ]\left (\mathcal{G}^{ -1}\delta \mathcal{G}\right )_{\mu _{4}}^{\sigma } -\frac{1}{2}W_{\mu _{1}\mu _{2}}^{\nu _{1}\nu _{2}}C_{\mu _{3}\mu _{4}}^{\nu _{3}}\left (\mathcal{G}^{ -1}\delta \mathcal{G}\right )\right \}\bigg\} .
\end{gather}
Finally, for the last term of \eqref{deltaIbulk}, we get
\begin{gather}
4\delta_{\nu_{1}\nu_{2}\nu_{3}}^{\mu_{1}\mu_{2}\mu_{3}}\delta\left(
C_{\mu_{1}}^{\nu_{1}\nu_{2}}C_{\mu_{2}\mu_{3}}^{\nu_{3}}\right)  =\delta
_{\nu_{1}\nu_{2}\nu_{3}}^{\mu_{1}\mu_{2}\mu_{3}}\Bigg\{\bigg\{-8\Big [\left(
\nabla^{\nu_{1}}S_{\mu_{1}}^{\sigma}\right)  C_{\mu_{2}\mu_{3}}^{\nu_{2}%
}+S_{\mu_{1}}^{\sigma}\left(  \nabla^{\nu_{1}}C_{\mu_{2}\mu_{3}}^{\nu_{2}%
}\right)  \Big ]\nonumber\\
-4\Big [\left(  \nabla_{\mu_{1}}S^{\sigma\nu_{1}}\right)  C_{\mu_{2}\mu_{3}%
}^{\nu_{2}}+S^{\sigma\nu_{1}}\left(  \nabla_{\mu_{1}}C_{\mu_{2}\mu_{3}}%
^{\nu_{2}}\right)  \Big ]+8\left(  \nabla^{\sigma}S_{\mu_{1}}^{\nu_{1}%
}\right)  C_{\mu_{2}\mu_{3}}^{\nu_{2}}+4\left[  \left(  \nabla_{\mu_{2}}%
C_{\mu_{1}}^{\nu_{1}\nu_{2}}\right)  S_{\mu_{3}}^{\sigma}-\frac{1}{2}%
C_{\mu_{1}}^{\nu_{1}\nu_{2}}C_{\mu_{2}\mu_{3}}^{\sigma}\right]  \nonumber\\
+2\left(  \nabla^{\nu_{1}}C_{\mu_{1}\mu_{2}}^{\nu_{2}}+\nabla_{\mu_{1}}%
C_{\mu_{2}}^{\nu_{1}\nu_{2}}\right)  R_{\mu_{3}}^{\sigma}-\nabla_{\mu_{3}%
}\nabla^{\sigma}\left(  \nabla^{\nu_{1}}C_{\mu_{1}\mu_{2}}^{\nu_{2}}%
+\nabla_{\mu_{1}}C_{\mu_{2}}^{\nu_{1}\nu_{2}}\right)  \bigg\}\left(
\mathcal{G}^{-1}\delta\mathcal{G}\right)  _{\sigma}^{\nu_{3}}\nonumber\\
+\bigg\{4\left[  -\left(  \nabla^{\nu_{1}}S_{\sigma}^{\nu_{2}}\right)
C_{\mu_{1}\mu_{2}}^{\nu_{3}}+S_{\sigma}^{\nu_{1}}\left(  \nabla^{\nu_{2}%
}C_{\mu_{1}\mu_{2}}^{\nu_{3}}\right)  \right]  +4\left[  \left(  \nabla
^{\nu_{3}}C_{\mu_{1}}^{\nu_{1}\nu_{2}}\right)  S_{\sigma\mu_{2}}+C_{\mu_{1}%
}^{\nu_{1}\nu_{2}}\left(  \nabla^{\nu_{3}}S_{\sigma\mu_{2}}\right)  \right]
\nonumber\\
-\nabla^{\nu_{3}}\nabla_{\sigma}\left(  \nabla^{\nu_{1}}C_{\mu_{1}\mu_{2}%
}^{\nu_{2}}+\nabla_{\mu_{1}}C_{\mu_{2}}^{\nu_{1}\nu_{2}}\right)
\bigg\}\left(  \mathcal{G}^{-1}\delta\mathcal{G}\right)  _{\mu_{3}}^{\sigma
}+\bigg\{4\left[  \left(  \nabla^{\sigma}S_{\sigma}^{\nu_{1}}\right)
C_{\mu_{1}\mu_{2}}^{\nu_{2}}+S_{\sigma}^{\nu_{1}}\left(  \nabla^{\sigma}%
C_{\mu_{1}\mu_{2}}^{\nu_{2}}\right)  \right]  \nonumber\\
-4\left[  \left(  \nabla_{\sigma}C_{\mu_{1}}^{\nu_{1}\nu_{2}}\right)
S_{\mu_{2}}^{\sigma}+C_{\mu_{1}}^{\nu_{1}\nu_{2}}\left(  \nabla_{\sigma}%
S_{\mu_{2}}^{\sigma}\right)  \right]  +\nabla^{\sigma}\nabla_{\sigma}\left(
\nabla^{\nu_{1}}C_{\mu_{1}\mu_{2}}^{\nu_{2}}+\nabla_{\mu_{1}}C_{\mu_{2}}%
^{\nu_{1}\nu_{2}}\right)  \bigg\}\left(  \mathcal{G}^{-1}\delta\mathcal{G}%
\right)  _{\mu_{3}}^{\nu_{3}}\nonumber\\
+\nabla^{\nu_{3}}\nabla_{\mu_{3}}\left(  \nabla^{\nu_{1}}C_{\mu_{1}\mu_{2}%
}^{\nu_{2}}+\nabla_{\mu_{1}}C_{\mu_{2}}^{\nu_{1}\nu_{2}}\right)  \left(
\mathcal{G}^{-1}\delta\mathcal{G}\right)  \nonumber\\
+\nabla^{\nu_{1}}\bigg[\left(  4S_{\sigma}^{\nu_{2}}C_{\mu_{1}\mu_{2}}%
^{\nu_{3}}-4C_{\mu_{1}}^{\nu_{2}\nu_{3}}S_{\sigma\mu_{2}}+\nabla_{\sigma
}\nabla^{\nu_{2}}C_{\mu_{1}\mu_{2}}^{\nu_{3}}+\nabla_{\sigma}\nabla_{\mu_{1}%
}C_{\mu_{2}}^{\nu_{2}\nu_{3}}\right)  \left(  \mathcal{G}^{-1}\delta
\mathcal{G}\right)  _{\mu_{3}}^{\sigma}\nonumber\\
+8S_{\mu_{1}}^{\sigma}C_{\mu_{2}\mu_{3}}^{\nu_{2}}\left(  \mathcal{G}%
^{-1}\delta\mathcal{G}\right)  _{\sigma}^{\nu_{3}}-\left(  \nabla_{\mu_{3}%
}\nabla^{\nu_{2}}C_{\mu_{1}\mu_{2}}^{\nu_{3}}+\nabla_{\mu_{3}}\nabla_{\mu_{1}%
}C_{\mu_{2}}^{\nu_{2}\nu_{3}}\right)  \left(  \mathcal{G}^{-1}\delta
\mathcal{G}\right)  +2C_{\mu_{1}\mu_{2}}^{\nu_{2}}\delta R_{\mu_{3}}^{\nu_{3}%
}\bigg]\nonumber\\
+\nabla_{\mu_{1}}\bigg[2C_{\mu_{2}}^{\nu_{1}\nu_{2}}\delta R_{\mu_{3}}%
^{\nu_{3}}+\left(  4S^{\sigma\nu_{1}}C_{\mu_{2}\mu_{3}}^{\nu_{2}}+4C_{\mu_{2}%
}^{\nu_{1}\nu_{2}}S_{\mu_{3}}^{\sigma}+\nabla^{\sigma}\nabla^{\nu_{1}}%
C_{\mu_{2}\mu_{3}}^{\nu_{2}}+\nabla^{\sigma}\nabla_{\mu_{2}}C_{\mu_{3}}%
^{\nu_{1}\nu_{2}}\right)  \left(  \mathcal{G}^{-1}\delta\mathcal{G}\right)
_{\sigma}^{\nu_{3}}\nonumber\\
+2\left(  \nabla^{\nu_{1}}C_{\mu_{2}\mu_{3}}^{\nu_{2}}+\nabla_{\mu_{2}}%
C_{\mu_{3}}^{\nu_{1}\nu_{2}}\right)  \mathcal{G}^{\kappa\nu_{3}}\delta
\Gamma_{\kappa\sigma}^{\sigma}\bigg]\nonumber\\
+\nabla_{\sigma}\bigg[\left(  4C_{\mu_{1}}^{\nu_{1}\nu_{2}}S_{\mu_{2}}%
^{\sigma}-4S^{\sigma\nu_{1}}C_{\mu_{1}\mu_{2}}^{\nu_{2}}-\nabla^{\sigma}%
\nabla^{\nu_{1}}C_{\mu_{1}\mu_{2}}^{\nu_{2}}-\nabla^{\sigma}\nabla_{\mu_{1}%
}C_{\mu_{2}}^{\nu_{1}\nu_{2}}\right)  \left(  \mathcal{G}^{-1}\delta
\mathcal{G}\right)  _{\mu_{3}}^{\nu_{3}}\nonumber\\
-2\left(  \nabla^{\nu_{1}}C_{\mu_{1}\mu_{2}}^{\nu_{2}}+\nabla_{\mu_{1}}%
C_{\mu_{2}}^{\nu_{1}\nu_{2}}\right)  \mathcal{G}^{\kappa\nu_{3}}\delta
\Gamma_{\kappa\mu_{3}}^{\sigma}\bigg]\Bigg \},
\end{gather}
where
\begin{equation}
2\delta _{\nu _{1}\nu _{2}\nu _{3}}^{\mu _{1}\mu _{2}\mu _{3}}C_{\mu _{2}}^{\nu _{1}\nu _{2}}\delta R_{\mu _{3}}^{\nu _{3}} = 2\delta _{\nu _{1}\nu _{2}\nu _{3}}^{\mu _{1}\mu _{2}\mu _{3}}C_{\mu _{2}}^{\nu _{1}\nu _{2}}\left [\mathcal{G}^{\sigma \nu _{3}}\left ( \nabla _{\lambda }\delta \Gamma _{\sigma \mu _{3}}^{\lambda } - \nabla _{\mu _{3}}\delta \Gamma _{\sigma \lambda }^{\lambda }\right ) -R_{\mu _{3}}^{\sigma }\left (\mathcal{G}^{ -1}\delta \mathcal{G}\right )_{\sigma }^{\nu _{3}}\right ] ,
\end{equation}
and

\begin{equation}
2\delta _{\nu _{1}\nu _{2}\nu _{3}}^{\mu _{1}\mu _{2}\mu _{3}}C_{\mu _{1}\mu _{2}}^{\nu _{2}}\delta R_{\mu _{3}}^{\nu _{3}} 
=2\delta _{\nu _{1}\nu _{2}\nu _{3}}^{\mu _{1}\mu _{2}\mu _{3}}C_{\mu _{1}\mu _{2}}^{\nu _{2}}\left [\mathcal{G}^{\sigma \nu _{3}}\left ( \nabla _{\lambda }\delta \Gamma _{\sigma \mu _{3}}^{\lambda } - \nabla _{\mu _{3}}\delta \Gamma _{\sigma \lambda }^{\lambda }\right ) -R_{\mu _{3}}^{\sigma }\left (\mathcal{G}^{ -1}\delta \mathcal{G} \right )_{\sigma }^{\nu _{3}}\right ] .
\end{equation}
At this moment we have in our hands the EOM of 6D CG for the Lu, Pang and Pope action. Summing up all the contributions one gets
\begin{gather*}
\delta I_{CG}^{\text{bulk}} =\int _{M}d^{6}x\sqrt{ -\mathcal{G}}\Bigg\{ -2\delta _{\nu _{1}\ldots \nu _{5}}^{\mu _{1}\ldots \mu _{5}}W_{\mu _{1}\mu _{2}}^{\nu _{1}\nu _{2}}S_{\mu _{3}}^{\nu _{3}}S_{\mu _{4}}^{\nu _{4}}\left (\mathcal{G}^{ -1}\delta \mathcal{G}\right )_{\mu _{5}}^{\nu _{5}} \\
+\delta _{\nu _{1}\ldots \nu _{4}}^{\mu _{1}\ldots \mu _{4}}\bigg\{2W_{\mu _{1}\mu _{2}}^{\nu _{1}\nu _{2}}S_{\mu _{3}}^{\nu _{3}}\left [2S_{\mu _{4}}^{\sigma }\left (\mathcal{G}^{ -1}\delta \mathcal{G}\right )_{\sigma }^{\nu _{4}} +S\left (\mathcal{G}^{ -1}\delta \mathcal{G}\right )_{\mu _{4}}^{\nu _{4}}\right ] -\frac{1}{2}W_{\mu _{1}\mu _{2}}^{\nu _{1}\nu _{2}}\left ( \nabla ^{\nu _{3}}C_{\mu _{3}\mu _{4}}^{\nu _{4}}\right )\left (\mathcal{G}^{ -1}\delta \mathcal{G}\right ) \\
-\left [\left ( \nabla ^{\nu _{4}} \nabla _{\sigma }W_{\mu _{1}\mu _{2}}^{\nu _{1}\nu _{2}}\right )S_{\mu _{3}}^{\nu _{3}} +\frac{1}{2}\left ( \nabla _{\sigma }W_{\mu _{1}\mu _{2}}^{\nu _{1}\nu _{2}}\right )C_{\mu _{3}}^{\nu _{3}\nu _{4}} +W_{\mu _{1}\mu _{2}}^{\nu _{1}\nu _{2}}\left ( \nabla ^{\nu _{4}} \nabla _{\sigma }S_{\mu _{3}}^{\nu _{3}}\right )\right ]\left (\mathcal{G}^{ -1}\delta \mathcal{G}\right )_{\mu _{4}}^{\sigma } \\
-\left [( \nabla _{\mu _{4}} \nabla ^{\sigma }W_{\mu _{1}\mu _{2}}^{\nu _{1}\nu _{2}})S_{\mu _{3}}^{\nu _{3}} +\frac{1}{2}( \nabla ^{\sigma }W_{\mu _{1}\mu _{2}}^{\nu _{1}\nu _{2}})C_{\mu _{3}\mu _{4}}^{\nu _{3}} +W_{\mu _{1}\mu _{2}}^{\nu _{1}\nu _{2}}\left ( \nabla _{\mu _{4}} \nabla ^{\sigma }S_{\mu _{3}}^{\nu _{3}}\right )\right ]\left (\mathcal{G}^{ -1}\delta \mathcal{G}\right )_{\sigma }^{\nu _{4}} +\Big [\left ( \nabla ^{\sigma } \nabla _{\sigma }W_{\mu _{1}\mu _{2}}^{\nu _{1}\nu _{2}}\right ) \\ 
+2\left ( \nabla ^{\sigma }W_{\mu _{1}\mu _{2}}^{\nu _{1}\nu _{2}}\right )\left ( \nabla _{\sigma }S_{\mu _{3}}^{\nu _{3}}\right ) +W_{\mu _{1}\mu _{2}}^{\nu _{1}\nu _{2}}\left ( \nabla ^{\sigma } \nabla _{\sigma }S_{\mu _{3}}^{\nu _{3}}\right )-4C_{\mu _{1}}^{\nu _{1}\nu _{2}}C_{\mu _{2}\mu _{3}}^{\nu _{3}} -8S_{\mu _{1}}^{\nu _{1}}\left ( \nabla ^{\nu _{2}}C_{\mu _{2}\mu _{3}}^{\nu _{3}}\right )\Big ]\left (\mathcal{G}^{ -1}\delta \mathcal{G}\right )_{\mu _{4}}^{\nu _{4}} \bigg\} \\
+\delta _{\nu _{1}\nu _{2}\nu _{3}}^{\mu _{1}\mu _{2}\mu _{3}}\bigg\{\bigg\{ -8\Big [\left ( \nabla ^{\nu _{1}}S_{\mu _{1}}^{\sigma }\right )C_{\mu _{2}\mu _{3}}^{\nu _{2}} +S_{\mu _{1}}^{\sigma }\left ( \nabla ^{\nu _{1}}C_{\mu _{2}\mu _{3}}^{\nu _{2}}\right )\Big ] -4\Big[\left ( \nabla _{\mu _{1}}S^{\sigma \nu _{1}}\right )C_{\mu _{2}\mu _{3}}^{\nu _{2}} +S^{\sigma \nu _{1}}\left ( \nabla _{\mu _{1}}C_{\mu _{2}\mu _{3}}^{\nu _{2}}\right )\Big ] \\
+14\left ( \nabla ^{\sigma }S_{\mu _{1}}^{\nu _{1}}\right )C_{\mu _{2}\mu _{3}}^{\nu _{2}}
+4\left [\left ( \nabla _{\mu _{2}}C_{\mu _{1}}^{\nu _{1}\nu _{2}}\right )S_{\mu _{3}}^{\sigma } -\frac{1}{2}C_{\mu _{1}}^{\nu _{1}\nu _{2}}C_{\mu _{2}\mu _{3}}^{\sigma }\right ] +2\left ( \nabla ^{\nu _{1}}C_{\mu _{1}\mu _{2}}^{\nu _{2}} + \nabla _{\mu _{1}}C_{\mu _{2}}^{\nu _{1}\nu _{2}}\right )R_{\mu _{3}}^{\sigma }  \\
-\nabla _{\mu _{3}} \nabla ^{\sigma }\left ( \nabla ^{\nu _{1}}C_{\mu _{1}\mu _{2}}^{\nu _{2}} + \nabla _{\mu _{1}}C_{\mu _{2}}^{\nu _{1}\nu _{2}}\right )\bigg\}\left (\mathcal{G}^{ -1}\delta \mathcal{G}\right )_{\sigma }^{\nu _{3}} 
+\bigg\{4\Big [ -\left ( \nabla ^{\nu _{1}}S_{\sigma }^{\nu _{2}}\right )C_{\mu _{1}\mu _{2}}^{\nu _{3}} +S_{\sigma }^{\nu _{1}}\left ( \nabla ^{\nu _{2}}C_{\mu _{1}\mu _{2}}^{\nu _{3}}\right )\Big ] \\
+4\Big [\left ( \nabla ^{\nu _{3}}C_{\mu _{1}}^{\nu _{1}\nu _{2}}\right )S_{\sigma \mu _{2}} +C_{\mu _{1}}^{\nu _{1}\nu _{2}}\left ( \nabla ^{\nu _{3}}S_{\sigma \mu _{2}}\right )\Big ] +6\left ( \nabla _{\sigma }S_{\mu _{1}}^{\nu _{1}}\right )C_{\mu _{2}}^{\nu _{2}\nu _{3}} - \nabla ^{\nu _{3}} \nabla _{\sigma }\left ( \nabla ^{\nu _{1}}C_{\mu _{1}\mu _{2}}^{\nu _{2}} + \nabla _{\mu _{1}}C_{\mu _{2}}^{\nu _{1}\nu _{2}}\right )\bigg\}\left (\mathcal{G}^{ -1}\delta \mathcal{G}\right )_{\mu _{3}}^{\sigma } \\ +\delta _{\nu _{1}\nu _{2}\nu _{3}}^{\mu _{1}\mu _{2}\mu _{3}}\bigg\{4\Big [\left ( \nabla ^{\sigma }S_{\sigma }^{\nu _{1}}\right )C_{\mu _{1}\mu _{2}}^{\nu _{2}} +S_{\sigma }^{\nu _{1}}\left ( \nabla ^{\sigma }C_{\mu _{1}\mu _{2}}^{\nu _{2}}\right )\Big ]-4\Big [\left ( \nabla _{\sigma }C_{\mu _{1}}^{\nu _{1}\nu _{2}}\right )S_{\mu _{2}}^{\sigma } +C_{\mu _{1}}^{\nu _{1}\nu _{2}}\left ( \nabla _{\sigma }S_{\mu _{2}}^{\sigma }\right )\Big ] + \\
\nabla ^{\sigma } \nabla _{\sigma }\left ( \nabla ^{\nu _{1}}C_{\mu _{1}\mu _{2}}^{\nu _{2}} + \nabla _{\mu _{1}}C_{\mu _{2}}^{\nu _{1}\nu _{2}}\right )\bigg\}\left (\mathcal{G}^{ -1}\delta \mathcal{G}\right )_{\mu _{3}}^{\nu _{3}}
+\Big [ \nabla ^{\nu _{3}} \nabla _{\mu _{3}}\left ( \nabla ^{\nu _{1}}C_{\mu _{1}\mu _{2}}^{\nu _{2}} + \nabla _{\mu _{1}}C_{\mu _{2}}^{\nu _{1}\nu _{2}}\right ) \\
+6S_{\mu _{1}}^{\nu _{1}}\left ( \nabla ^{\nu _{2}}C_{\mu _{2}\mu _{3}}^{\nu _{3}}\right ) +8C_{\mu _{1}}^{\nu _{1}\nu _{2}}C_{\mu _{2}\mu _{3}}^{\nu _{3}}\Big ]\left (\mathcal{G}^{ -1}\delta \mathcal{G}\right )\bigg\}\Bigg\} +b .t .
\end{gather*}
Here, the bulk term provides the EOM of the theory in a more compact form than the one given by Lu, Pang and Pope. Thus, the obstruction tensor that defines the EOM of LPP CG reads
\begin{gather*}
E_{\nu }^{\mu } = -2\delta _{\nu _{1}\ldots \nu _{4}\nu }^{\mu _{1}\ldots \mu _{4}\mu }W_{\mu _{1}\mu _{2}}^{\nu _{1}\nu _{2}}S_{\mu _{3}}^{\nu _{3}}S_{\mu _{4}}^{\nu _{4}}  
-\frac{1}{2}\delta _{\nu _{1}\ldots \nu _{4}}^{\mu _{1}\ldots \mu _{4}}W_{\mu _{1}\mu _{2}}^{\nu _{1}\nu _{2}}\left ( \nabla ^{\nu _{3}}C_{\mu _{3}\mu _{4}}^{\nu _{4}}\right )\delta _{\nu }^{\mu } \\
-\delta _{\nu _{1}\ldots \nu _{4}}^{\mu _{1}\ldots \mu _{3}\mu }\left [\left ( \nabla ^{\nu _{4}} \nabla _{\nu }W_{\mu _{1}\mu _{2}}^{\nu _{1}\nu _{2}}\right )S_{\mu _{3}}^{\nu _{3}} +\frac{1}{2}\left ( \nabla _{\nu }W_{\mu _{1}\mu _{2}}^{\nu _{1}\nu _{2}}\right )C_{\mu _{3}}^{\nu _{3}\nu _{4}} +W_{\mu _{1}\mu _{2}}^{\nu _{1}\nu _{2}}\left ( \nabla ^{\nu _{4}} \nabla _{\nu }S_{\mu _{3}}^{\nu _{3}}\right )\right ] \\
-\delta _{\nu _{1}\ldots \nu _{3}\nu }^{\mu _{1}\ldots \mu _{4}}\left [( \nabla _{\mu _{4}} \nabla ^{\mu }W_{\mu _{1}\mu _{2}}^{\nu _{1}\nu _{2}})S_{\mu _{3}}^{\nu _{3}} +\frac{1}{2}( \nabla ^{\mu }W_{\mu _{1}\mu _{2}}^{\nu _{1}\nu _{2}})C_{\mu _{3}\mu _{4}}^{\nu _{3}} +W_{\mu _{1}\mu _{2}}^{\nu _{1}\nu _{2}}\left ( \nabla _{\mu _{4}} \nabla ^{\mu }S_{\mu _{3}}^{\nu _{3}}\right ) -4W_{\mu _{1}\mu _{2}}^{\nu _{1}\nu _{2}}S_{\mu _{3}}^{\nu _{3}}S_{\mu _{4}}^{\mu }\right ] \\
+\delta _{\nu _{1}\ldots \nu _{3}\nu}^{\mu _{1}\ldots \mu _{3}\mu } \Big[ \left ( \nabla ^{\sigma } \nabla _{\sigma }W_{\mu _{1}\mu _{2}}^{\nu _{1}\nu _{2}}\right ) +2\left ( \nabla ^{\sigma }W_{\mu _{1}\mu _{2}}^{\nu _{1}\nu _{2}}\right )\left ( \nabla _{\sigma }S_{\mu _{3}}^{\nu _{3}}\right ) +W_{\mu _{1}\mu _{2}}^{\nu _{1}\nu _{2}}\left ( \nabla ^{\sigma } \nabla _{\sigma }S_{\mu _{3}}^{\nu _{3}}\right ) -  4 C_{\mu _{1}}^{\nu _{1}\nu _{2}}C_{\mu _{2}\mu _{3}}^{\nu _{3}} \\
-8 S_{\mu _{1}}^{\nu _{1}}\left ( \nabla ^{\nu _{2}}C_{\mu _{2}\mu _{3}}^{\nu _{3}}\right ) +2 S W_{\mu _{1}\mu _{2}}^{\nu _{1}\nu _{2}}S_{\mu _{3}}^{\nu _{3}} \Big]
+\delta _{\nu _{1}\nu _{2}\nu }^{\mu _{1}\mu _{2}\mu _{3}}\bigg\{ -8 \nabla ^{\nu _{1}}\left (S_{\mu _{1}}^{\mu }C_{\mu _{2}\mu _{3}}^{\nu _{2}}\right ) -4 \nabla _{\mu _{1}}\left (S^{\mu \nu _{1}}C_{\mu _{2}\mu _{3}}^{\nu _{2}}\right ) +14\left ( \nabla ^{\mu }S_{\mu _{1}}^{\nu _{1}}\right )C_{\mu _{2}\mu _{3}}^{\nu _{2}} \\ 
+2\left ( \nabla ^{\nu _{1}}C_{\mu _{1}\mu _{2}}^{\nu _{2}} + \nabla _{\mu _{1}}C_{\mu _{2}}^{\nu _{1}\nu _{2}}\right )R_{\mu _{3}}^{\mu }
+4\left [\left ( \nabla _{\mu _{2}}C_{\mu _{1}}^{\nu _{1}\nu _{2}}\right )S_{\mu _{3}}^{\mu } -\frac{1}{2}C_{\mu _{1}}^{\nu _{1}\nu _{2}}C_{\mu _{2}\mu _{3}}^{\mu }\right ] - \nabla _{\mu _{3}} \nabla ^{\mu }\left ( \nabla ^{\nu _{1}}C_{\mu _{1}\mu _{2}}^{\nu _{2}} + \nabla _{\mu _{1}}C_{\mu _{2}}^{\nu _{1}\nu _{2}}\right )\bigg\} \\
+\delta _{\nu _{1}\nu _{2}\nu _{3}}^{\mu _{1}\mu _{2}\mu } \Big[ -4 \nabla ^{\nu _{1}}\left (S_{\nu }^{\nu _{2}}C_{\mu _{1}\mu _{2}}^{\nu _{3}}\right ) +4 \nabla ^{\nu _{3}}\left (C_{\mu _{1}}^{\nu _{1}\nu _{2}}S_{\nu \mu _{2}}\right ) +6\left ( \nabla _{\nu }S_{\mu _{1}}^{\nu _{1}}\right )C_{\mu _{2}}^{\nu _{2}\nu _{3}} 
- \nabla ^{\nu _{3}} \nabla _{\nu }\left ( \nabla ^{\nu _{1}}C_{\mu _{1}\mu _{2}}^{\nu _{2}} + \nabla _{\mu _{1}}C_{\mu _{2}}^{\nu _{1}\nu _{2}}\right ) \Big] \\ +\delta _{\nu _{1}\nu _{2}\nu}^{\mu _{1}\mu _{2}\mu} \Big[4 \nabla ^{\sigma }\left (S_{\sigma }^{\nu _{1}}C_{\mu _{1}\mu _{2}}^{\nu _{2}}\right ) 
-4 \nabla _{\sigma }\left (C_{\mu _{1}}^{\nu _{1}\nu _{2}}S_{\mu _{2}}^{\sigma }\right ) + \nabla ^{\sigma } \nabla _{\sigma }\left ( \nabla ^{\nu _{1}}C_{\mu _{1}\mu _{2}}^{\nu _{2}} + \nabla _{\mu _{1}}C_{\mu _{2}}^{\nu _{1}\nu _{2}}\right ) \Big] \\
+\delta _{\nu _{1}\nu _{2}\nu _{3}}^{\mu _{1}\mu _{2}\mu _{3}} \Big[ \nabla ^{\nu _{3}} \nabla _{\mu _{3}}\left ( \nabla ^{\nu _{1}}C_{\mu _{1}\mu _{2}}^{\nu _{2}} + \nabla _{\mu _{1}}C_{\mu _{2}}^{\nu _{1}\nu _{2}}\right ) +6S_{\mu _{1}}^{\nu _{1}}\left ( \nabla ^{\nu _{2}}C_{\mu _{2}\mu _{3}}^{\nu _{3}}\right ) +8C_{\mu _{1}}^{\nu _{1}\nu _{2}}C_{\mu _{2}\mu _{3}}^{\nu _{3}} \Big] \delta _{\nu }^{\mu } .
\end{gather*}

\bibliography{6DCGv6bib}
\bibliographystyle{JHEP-2}

\end{document}